\documentclass[preprint,12pt]{elsarticle}

\usepackage{amssymb}

\usepackage{amsmath}

\usepackage{subfig}

\usepackage{booktabs}

\usepackage{textcomp}

\journal{Information Fusion}

\begin{document}

\begin{frontmatter}



\title{Applying Gaussian distributed constraints to Gaussian distributed variables}


\author{Andrew W. Palmer, Andrew J. Hill, Steven J. Scheding \corref{cor1}}
\cortext[cor1]{Email address: \{a.palmer;a.hill;s.scheding\}@acfr.usyd.edu.au}

\address{Australian Centre for Field Robotics, The University of Sydney, NSW, Australia}

\begin{abstract}

This paper develops an analytical method of truncating inequality constrained Gaussian distributed variables where the constraints are themselves described by Gaussian distributions. Existing truncation methods either assume hard constraints, or use numerical methods to handle uncertain constraints. The proposed approach introduces moment-based Gaussian approximations of the truncated distribution. This method can be applied to numerous problems, with the motivating problem being Kalman filtering with uncertain constraints. In a simulation example, the developed method is shown to outperform unconstrained Kalman filtering by over 40\% and hard-constrained Kalman filtering by over 17\%.
\footnote{\textcopyright 2016. This manuscript version is made available under the CC-BY-NC-ND 4.0 license http://creativecommons.org/licenses/by-nc-nd/4.0/} 

\end{abstract}


\end{frontmatter}


\section{Introduction}

Gaussian distributed variables are widely used to represent the state of a system in many problems ranging from state estimation \cite{Simon2006b} to scheduling \cite{Palmer2013,Palmer2014a}. In practice, the state vectors in many systems are known to satisfy inequality constraints. Examples of state-constrained systems include health monitoring \cite{Simon2006}, vision systems \cite{Shimada1998}, robotics \cite{Boccadoro2010}, binary sensor networks \cite{Manes2013}, and object tracking \cite{Romero-cano2015}. This paper deals specifically with systems that are subject to inequality constraints where the constraints themselves have uncertainty described by Gaussian distributions. Constraints described by Gaussian distributions can arise from many sources in state estimation problems including discrete sensors, such as position or level switches, that have uncertainty on their activation point, obstacles whose positions are uncertain, and other physical and model-derived bounds such as maximum fuel levels based on historical fuel burn rates. Constrained Gaussian distributed variables also appear in scheduling applications where the distribution describing the time at which an event is predicted to occur is constrained by the time distributions of other events. 

\begin{figure}
\centering
\includegraphics[width = 0.8\textwidth]{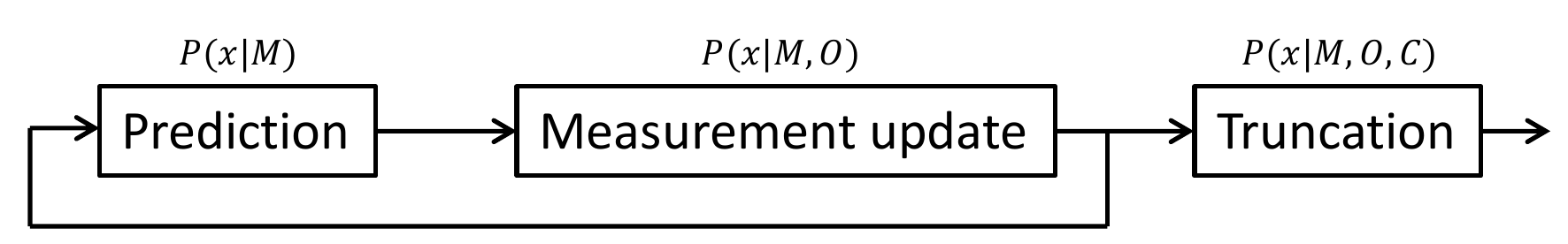}
\caption{The Kalman filter is run independent of the truncation method, with the truncation being applied to the state estimate that is the output of the Kalman filter. The prediction step of the Kalman filter results in a probability distribution describing the state, $x$, conditioned on the system model, $M$. The measurement update step further conditions the state estimate on the observations, $O$. Finally, the truncation step conditions the estimate on the constraints acting on the state, $C$. } \label{f:truncation}
\end{figure}

Hard inequality constraints are well studied \cite{Simon2006b}, where the main approaches are estimate projection \cite{Simon2006}, gain projection \cite{Gupta2007}, and Probability Density Function (PDF) truncation \cite{Simon2010b}. Estimate and gain projection approaches incorporate the constraints into the derivation of the Kalman filter, resulting in a constrained optimisation problem that can be solved using quadratic programming, least squares approaches, amongst others \cite{Simon2006b, Simon2010}. Truncation methods, on the other hand, are applied directly to the PDF resulting from a Kalman filter, as outlined in Figure \ref{f:truncation}. This approach truncates the PDF at the constraints and calculates the mean and covariance of the truncated PDF, which become the constrained state estimate and its covariance. The PDF truncation approach was shown in \cite{Simon2010b} to, in general, outperform the estimate projection method. The truncation approach has been applied to probabilistic collision checking for robots \cite{Patil2012}, and has been extended to non-linear systems \cite{Teixeira2010,Straka2012}. 

Soft constraints correspond to uncertain or noisy constraints, and are less studied than hard constraints. Soft equality constraints are typically incorporated as noisy measurements \cite{Simon2006b,Helor1993}. However, soft inequality constraints are significantly more difficult to deal with, and numerical filters such as a Particle Filter (PF) are typically used for these problems \cite{Shao2010}. Several numerical methods have been examined for incorporating soft constraints into the Kalman filter. A numerical PDF truncation method was used in \cite{Boccadoro2010} for robot localisation using Radio Frequency IDentification (RFID) tags, where the noise on the inequality constraints was highly non-Gaussian. Compared with a PF approach, the numerical PDF truncation method was 2 to 3 orders of magnitude faster while, in general, providing similar results. A similar RFID problem was examined in \cite{Manes2013} where aspects of the Unscented Kalman Filter (UKF) and PF were combined---the prediction step used the standard UKF step, while the correction step was modified to weight the sigma-points of the UKF in a similar manner to the weighting process in a PF. It was shown to outperform a PF as well as the Quantised Extended Kalman Filter (QEKF) presented in \cite{DiGiampaolo2012}.

The literature on soft inequality constraints has focused on constraints with non-Gaussian distributions, where the constrained state estimates are, by necessity, calculated using numerical methods. The main contribution of this paper is an analytical method for PDF truncation with soft constraints where the soft constraints are described by Gaussian distributions. This reduces the computational requirement compared to numerical methods, and it is shown to provide superior estimation performance compared to unconstrained and hard-constrained state estimation methods. The truncation approach presented in this paper is not limited to Kalman filters and can be applied to any constrained system using Gaussian distributions to represent the state and constraints. 

The rest of this paper is structured as follows: Section \ref{s:probdef} introduces the constrained Kalman filtering problem, Section \ref{s:transform} shows how the state and constraints can be transformed such that each state has only one constraint acting on it, Section \ref{s:constrained} presents the truncation method for a one-sided constraint, and Section \ref{s:interval} extends this to an interval constraint. The performance of the methods are evaluated in Section \ref{s:results}, and the paper is concluded in Section \ref{s:conc}. \ref{s:a_mean} and \ref{s:a_variance} provide in-depth derivations of the integrals used in this paper. 

\section{Problem definition}\label{s:probdef}

This paper adapts the notation used in \cite{Simon2010b}. A discrete linear time-invariant system is described by:

\begin{gather}
\boldsymbol{x}\left(k\right) = \boldsymbol{Fx}\left(k-1\right) + \boldsymbol{Gu}\left(k\right) + \boldsymbol{w}\left(k\right) \notag \\
\boldsymbol{y}\left(k\right) = \boldsymbol{Hx}\left(k\right) + \boldsymbol{v}\left(k\right)
\end{gather}
where $k$ is the time index, $\boldsymbol{x}$ is the state vector with $n$ states, $\boldsymbol{u}$ is the vector of known control inputs, and $\boldsymbol{y}$ is the vector of measurements. The vectors $\boldsymbol{w}$ and $\boldsymbol{v}$ contain the process and measurement noise respectively. The process noise, $\boldsymbol{w}$, is assumed to be zero mean Gaussian white noise with a covariance matrix of $\boldsymbol{Q}$. The measurement noise, $\boldsymbol{v}$, is similarly assumed to be zero mean Gaussian white noise with a covariance matrix of $\boldsymbol{R}$. The noises at each time-step are assumed to be independent. 

For the given system, the Kalman filter prediction equations are \cite{Faragher2012}:

\begin{gather}
\boldsymbol{\hat{x}}(k|k-1) = \boldsymbol{F\hat{x}}(k-1|k-1) + \boldsymbol{Gu}(k-1) \notag \\
\boldsymbol{P}(k|k-1) = \boldsymbol{FP}(k-1|k-1)\boldsymbol{F}^{T} + \boldsymbol{Q}
\end{gather}
and the measurement update equations are:

\begin{gather}
\boldsymbol{K}= \boldsymbol{P}(k|k-1) \boldsymbol{H}^{T} \left( \boldsymbol{HP}(k|k-1)\boldsymbol{H}^{T} + \boldsymbol{R} \right)^{-1} \notag \\
\boldsymbol{\hat{x}}(k|k) = \boldsymbol{\hat{x}}(k|k-1) + \boldsymbol{K} \left( \boldsymbol{y}(k) - \boldsymbol{H\hat{x}}(k|k-1)  \right) \\
\boldsymbol{P}(k|k) = \boldsymbol{P}(k|k-1) - \boldsymbol{K}\boldsymbol{H}\boldsymbol{P}(k|k-1) \notag
\end{gather}
where $\boldsymbol{\hat{x}}(k|k)$ is the state estimate, and $\boldsymbol{P}(k|k)$ is the covariance of the state estimate. The state estimate is initialised with $\boldsymbol{\hat{x}}(0) = E[\boldsymbol{x}(0)]$, where $E[.]$ is the expectation operator. The covariance matrix is initialised with $\boldsymbol{P}(0) = E[(\boldsymbol{x}(0) - \boldsymbol{\hat{x}}(0))(\boldsymbol{x}(0) - \boldsymbol{\hat{x}}(0))^{T}]$. 

Now consider the following $s$ linearly independent constraints on the system:
\begin{equation}\label{eq:constraint_def}
A_{m}(k) \le \boldsymbol{\phi}_{m}^{T}(k)\boldsymbol{x}(k) \le B_{m}(k) \qquad m=1,...,s
\end{equation}
where the constraints are uncertain and normally distributed:
\begin{equation}
A_{m}(k) \sim \mathcal{N}(\mu_{a,m},\sigma_{a,m}^{2}) \qquad B_{m}(k) \sim \mathcal{N}(\mu_{b,m},\sigma_{b,m}^{2})
\end{equation}

Equation (\ref{eq:constraint_def}) describes a two-sided constraint on the linear function of the state described by $\boldsymbol{\phi}_{m}^{T}(k)\boldsymbol{x}(k)$. One sided constraints can be represented by setting $\mu_{a,m} = -\infty$, or $\mu_{b,m} = \infty$, and hard constraints can be implemented by setting $\sigma_{a,m} \approx 0$ or $\sigma_{b,m} \approx 0$ as required. 

Given an estimate $\boldsymbol{\hat{x}}(k)$ with covariance $\boldsymbol{P}(k)$ at time $k$, the problem is to truncate the Gaussian PDF $\mathcal{N}(\boldsymbol{\hat{x}}(k),\boldsymbol{P}(k))$ using the $s$ constraints described above, and then find the mean $\boldsymbol{\tilde{x}}(k)$ and covariance $\boldsymbol{\tilde{P}}(k)$ of the truncated PDF. The calculated mean and covariance represent the constrained estimate of the state. 

\section{Transforming the state vector and constraints}\label{s:transform}

To apply the constraints via the truncation method, the state vector must be transformed so that the constraints are decoupled. This will result in $s$ transformed constraints that each involve only one element of the transformed state, allowing the constraints to be enforced individually on each element of the transformed state. It should be noted that the order in which constraints are applied can change the final state estimate. However, if the initial constraints are decoupled, the order of constraint application does not change the result \cite{Simon2010b}. 

The transformation process is outlined in \cite{Simon2006b} and \cite{Simon2010b}, and is summarised here in equations (\ref{eq:transform_start})--(\ref{eq:transform_end}) and (\ref{eq:mean_and_sigma})--(\ref{eq:truncated_estimate}). For ease of notation, the $(k)$ after each variable will be dropped. Let the vector $\boldsymbol{\tilde{x}}_{i}$ be the truncated state estimate, and the matrix $\boldsymbol{\tilde{P}}_{i}$ be the covariance of $\boldsymbol{\tilde{x}}_{i}$, after the first $i-1$ constraints have been enforced. To initialise the process:

\begin{gather}
i = 1 \quad \boldsymbol{\tilde{x}}_{i} = \boldsymbol{\hat{x}} \quad \boldsymbol{\tilde{P}}_{i} = \boldsymbol{{P}} \label{eq:transform_start}
\end{gather}

The transformed state vector is given by:

\begin{equation} \label{eq:transform}
\boldsymbol{z}_{i} = \boldsymbol{\rho}_{i}\boldsymbol{W}_{i}^{-1/2}\boldsymbol{T}_{i}^{T}(\boldsymbol{x}-\boldsymbol{\tilde{x}}_{i})
\end{equation}
where the matrices $\boldsymbol{T}_{i}$ and $\boldsymbol{W}_{i}$ are derived from the Jordan canonical decomposition of $\boldsymbol{\tilde{P}}_{i}$:

\begin{equation}
\boldsymbol{T}_{i}\boldsymbol{W}_{i}\boldsymbol{T}_{i}^{T} = \boldsymbol{\tilde{P}}_{i}
\end{equation}

$\boldsymbol{T}_{i}$ is an orthogonal matrix, and $\boldsymbol{W}_{i}$ is a diagonal matrix. The matrix $\boldsymbol{\rho}_{i}$ is derived by the Gram--Schmidt orthogonalisation \cite{Moon2000} which finds the orthogonal $\boldsymbol{\rho}_{i}$ that satisfies:

\begin{equation}
\boldsymbol{\rho}_{i}\boldsymbol{W}_{i}^{1/2}\boldsymbol{T}_{i}^{T}\boldsymbol{\phi}_{i} = \left[(\boldsymbol{\phi}_{i}^{T}\boldsymbol{\tilde{P}}_{i}\boldsymbol{\phi}_{i})^{1/2} \quad 0 \quad ... \quad 0 \right]^{T}
\end{equation}

Now only one element of $\boldsymbol{z}_{i}$ is constrained, and the states in the transformed state vector $\boldsymbol{z}_{i}$ are independent standard normal distributions. Let $\boldsymbol{e}_{i}$ be the $i$th column of an $n \times n$ identity matrix. Transforming the constraints results in:

\begin{equation}
C_{i} \le \boldsymbol{e}_{i}^{T}\boldsymbol{z}_{i} \le D_{i}
\end{equation}
where
\begin{gather}
C_{i} \sim \mathcal{N}(\mu_{c,i},\sigma_{c,i}^{2}) \notag \\
\mu_{c,i} = \frac{\mu_{a,i} - \boldsymbol{\phi}_{i}^{T}\boldsymbol{\tilde{x}}_{i}}{\sqrt{\boldsymbol{\phi}_{i}^{T}\boldsymbol{\tilde{P}}_{i}\boldsymbol{\phi}_{i}}} \quad
\sigma_{c,i} = \frac{\sigma_{a,i}}{\sqrt{\boldsymbol{\phi}_{i}^{T}\boldsymbol{\tilde{P}}_{i}\boldsymbol{\phi}_{i}}}
\end{gather}
and 
\begin{gather}
D_{i} \sim \mathcal{N}(\mu_{d,i},\sigma_{d,i}^{2}) \notag \\
\mu_{d,i} = \frac{\mu_{b,i} - \boldsymbol{\phi}_{i}^{T}\boldsymbol{\tilde{x}}_{i}}{\sqrt{\boldsymbol{\phi}_{i}^{T}\boldsymbol{\tilde{P}}_{i}\boldsymbol{\phi}_{i}}} \quad
\sigma_{d,i} = \frac{\sigma_{b,i}}{\sqrt{\boldsymbol{\phi}_{i}^{T}\boldsymbol{\tilde{P}}_{i}\boldsymbol{\phi}_{i}}} \label{eq:transform_end}
\end{gather}

The equations for calculating the standard deviation of each constraint are not present in \cite{Simon2006b,Simon2010b}, but they are a trivial extension from the equations provided for calculating the mean. 

\section{One-sided constraint}\label{s:constrained}

First, consider the case where there is only one constraint on the transformed state, in this case a lower constraint:

\begin{equation}
C_{i} \le \boldsymbol{e}_{i}^{T}\boldsymbol{z}_{i} \label{eq:singleconstraint}
\end{equation} 

Applying a lower constraint to the transformed state is equivalent to finding the conditional probability distribution of the transformed state given that it is higher than the constraint. Using Bayes' theorem, the conditional probability distribution, $p_{\boldsymbol{e}_{i}^{T}\boldsymbol{z}_{i}}(\zeta | C_{i} \le \zeta)$, as a function of $\zeta$ is given by:

\begin{equation} \label{eq:bayes_single}
p_{\boldsymbol{e}_{i}^{T}\boldsymbol{z}_{i}}(\zeta | C_{i} \le \zeta) = \frac{p_{\boldsymbol{e}_{i}^{T}\boldsymbol{z}_{i}}(\zeta) \times P(C_{i} \le \zeta)} {P(C_{i} \le \boldsymbol{e}_{i}^{T}\boldsymbol{z}_{i})}
\end{equation}
where $p_{\boldsymbol{e}_{i}^{T}\boldsymbol{z}_{i}}(\zeta)$ is the PDF of $\boldsymbol{e}_{i}^{T}\boldsymbol{z}_{i}$, $P(C_{i} \le \zeta)$ is the probability that a point $\zeta$ is greater than the constraint, and $P(C_{i} \le \boldsymbol{e}_{i}^{T}\boldsymbol{z}_{i})$ is the probability that the transformed state is greater than the constraint. $P(C_{i} \le \zeta)$ is given by:

\begin{align}
P(C_{i} \le \zeta) &= \int\limits_{-\infty}^{\zeta} \textrm{PDF}_{C_{i}}(c) \; \textrm{d}c  \notag  \\
& = \textrm{CDF}_{C_{i}}(\zeta)
\end{align}
where $\textrm{PDF}_{C_{i}}(c)$ is the PDF of the constraint $C_{i}$ evaluated at $c$, and $\textrm{CDF}_{C_{i}}(\zeta)$ is the Cumulative Distribution Function (CDF) of $C_{i}$ evaluated at $\zeta$. $P(C_{i} \le \boldsymbol{e}_{i}^{T}\boldsymbol{z}_{i})$ is given by:

\begin{align}
P(C_{i} \le \boldsymbol{e}_{i}^{T}\boldsymbol{z}_{i}) & = P(C_{i} - \boldsymbol{e}_{i}^{T}\boldsymbol{z}_{i} \le 0) \notag \\
 & = \int\limits_{-\infty}^{0}\textrm{PDF}_{C_{i} - \boldsymbol{e}_{i}^{T}\boldsymbol{z}_{i}}(\zeta) \; \textrm{d}\zeta \notag \\
 & = \textrm{CDF}_{C_{i} - \boldsymbol{e}_{i}^{T}\boldsymbol{z}_{i}}(0)
\end{align}
where $C_{i} - \boldsymbol{e}_{i}^{T}\boldsymbol{z}_{i} \sim \mathcal{N}(\mu_{c,i},\sigma_{c,i}^{2} + 1)$ since $\boldsymbol{e}_{i}^{T}\boldsymbol{z}_{i}$ is a standard normal distribution. The conditional probability distribution of the transformed state given that it is higher than the constraint is then given by:

\begin{align}\label{eq:bayes_single_pdfs}
p_{\boldsymbol{e}_{i}^{T}\boldsymbol{z}_{i}}(\zeta | C_{i} \le \zeta) &= \frac{p_{\boldsymbol{e}_{i}^{T}\boldsymbol{z}_{i}}(\zeta) \times P(C_{i} \le \zeta)} {P(C_{i} \le \boldsymbol{e}_{i}^{T}\boldsymbol{z}_{i})} \notag \\
& = \frac{\textrm{PDF}_{\boldsymbol{e}_{i}^{T}\boldsymbol{z}_{i}}(\zeta) \times \textrm{CDF}_{C_{i}}(\zeta)}{\textrm{CDF}_{C_{i} - \boldsymbol{e}_{i}^{T}\boldsymbol{z}_{i}}(0)}
\end{align}

The denominator of (\ref{eq:bayes_single_pdfs}) can be thought of as a normalising factor---it is the area of the numerator and ensures that the CDF of $p_{\boldsymbol{e}_{i}^{T}\boldsymbol{z}_{i}}(\zeta | C_{i} \le \zeta)$ is bound between 0 and 1. For states and constraints described by Gaussian distributions, $p_{\boldsymbol{e}_{i}^{T}\boldsymbol{z}_{i}}(\zeta | C_{i} \le \zeta)$ is given by:

\begin{equation} \label{eq:dist_single}
p_{\boldsymbol{e}_{i}^{T}\boldsymbol{z}_{i}}(\zeta | C_{i} \le \zeta) = \frac{\frac{1}{\sqrt{2\pi}}\exp(\frac{-x^{2}}{2})\frac{1}{2}\left[1 + \textrm{erf}\left( \frac{\zeta - \mu_{c,i}}{\sigma_{c,i}\sqrt{2}} \right) \right]}{\frac{1}{2}\left[1-\textrm{erf}\left(\frac{\mu_{c,i}}{\sqrt{2\left(\sigma_{c,i}^{2} + 1\right)}}\right)\right]}
\end{equation}
where erf(.) is the error function, defined as:

\begin{equation}
\textrm{erf}(t) = \frac{2}{\sqrt{\pi}}\int\limits_{0}^{t}\exp\left(-\tau^{2}\right)\textrm{d}\tau
\end{equation}

Let:

\begin{equation}
\alpha_{i} = \frac{1}{\sqrt{2\pi}\left[1-\textrm{erf}\left(\frac{\mu_{c,i}}{\sqrt{2\left(\sigma_{c,i}^{2} + 1\right)}}\right)\right]}
\end{equation}
then
\begin{equation} \label{eq:dist_single_simp}
p_{\boldsymbol{e}_{i}^{T}\boldsymbol{z}_{i}}(\zeta | C_{i} \le \zeta) = \alpha_{i}\exp\left(-\zeta^{2}/2\right)\left[1+\textrm{erf}\left(\frac{\zeta-\mu_{c,i}}{\sigma_{c,i}\sqrt{2}} \right)\right]
\end{equation}

To approximate $p_{\boldsymbol{e}_{i}^{T}\boldsymbol{z}_{i}}(\zeta | C_{i} \le \zeta)$ with a Gaussian distribution, the mean and variance are calculated as follows:

\begin{align} 
\mu_{i} &= \alpha_{i}\int\limits_{-\infty}^{\infty}\zeta \exp\left(-\zeta^{2}/2\right)\left[1+\textrm{erf}\left(\frac{\zeta-\mu_{c,i}}{\sigma_{c,i}\sqrt{2}} \right)\right] \textrm{d}\zeta    \notag\\
&= \frac{2\alpha_{i}}{\sqrt{\sigma_{c,i}^2 + 1}}\exp\left(-\frac{\mu_{c,i}^{2}}{2(\sigma_{c,i}^2 + 1)}\right) \label{eq:single_mu}
\end{align}

\begin{equation} \label{eq:single_sigma}
\begin{split}
\sigma_{i}^{2} &= \alpha_{i}\int\limits_{-\infty}^{\infty}\left(\zeta - \mu_{i}\right)^{2} \exp\left(-\zeta^{2}/2\right)\left[1+\textrm{erf}\left(\frac{\zeta-\mu_{c,i}}{\sigma_{c,i}\sqrt{2}} \right)\right] \textrm{d}\zeta    \\
&= \alpha_{i}\left[\sqrt{2\pi}\left(\left(1+\mu_{i}^{2}\right)\left(1-\textrm{erf}\left(\frac{\mu_{c,i}}{\sqrt{2(\sigma_{c,i}^{2}+1)}}\right)\right)\right) \right. \\
& \qquad + \left. \frac{2}{\sqrt{\sigma_{c,i}^2 + 1}}\exp\left(-\frac{\mu_{c,i}^{2}}{2(\sigma_{c,i}^2 + 1)}\right)\left(\frac{\mu_{c,i}}{\sigma_{c,i}^2 + 1} - 2\mu_{i} \right)\right]
\end{split}
\end{equation}

The derivations of the mean and variance can be found in \ref{s:a_mean} and \ref{s:a_variance} respectively. The transformed state estimate, after the $i$th constraint has been applied, has the following mean and covariance:

\begin{gather}
\boldsymbol{\tilde{z}}_{i+1} = \mu_{i}\boldsymbol{e}_{i}\notag \\
\boldsymbol{\tilde{G}}_{i+1} = \boldsymbol{I}_{n} + \left(\sigma_{i}^{2}-1\right)\boldsymbol{e}_{i}\boldsymbol{e}_{i}^{T}   \label{eq:mean_and_sigma}
\end{gather}
where $\boldsymbol{I}_{n}$ is an $n \times n$ identity matrix. Taking the inverse of the transformation in (\ref{eq:transform}) gives the mean and variance of the state estimate after the truncation of the $i$th constraint:

\begin{gather} 
\boldsymbol{\tilde{x}}_{i+1} = \boldsymbol{T}_{i}\boldsymbol{W}_{i}^{1/2}\boldsymbol{\rho}_{i}^{T}\boldsymbol{\tilde{z}}_{i+1}+\boldsymbol{\tilde{x}}_{i} \notag \\
\boldsymbol{\tilde{P}}_{i+1} = \boldsymbol{T}_{i}\boldsymbol{W}_{i}^{1/2}\boldsymbol{\rho}_{i}^{T}\boldsymbol{\tilde{G}}_{i+1}\boldsymbol{\rho}_{i}\boldsymbol{W}_{i}^{1/2}\boldsymbol{T}_{i}^{T} \label{eq:inverse_transform}
\end{gather}

This process (from (\ref{eq:transform}) to (\ref{eq:inverse_transform})) is repeated for the $s$ constraints, incrementing $i$ each time and using the constrained state estimate after constraint $i$ has been applied as the input state estimate for constraint $i+1$. After the $s$ constraints have been applied, the constrained state estimate is:

\begin{gather}
\boldsymbol{\tilde{x}} = \boldsymbol{\tilde{x}}_{s+1} \notag \\
\boldsymbol{\tilde{P}} = \boldsymbol{\tilde{P}}_{s+1} \label{eq:truncated_estimate}
\end{gather}

The equations for applying an upper constraint of the form:

\begin{equation}
\boldsymbol{e}_{i}^{T}\boldsymbol{z}_{i} \le D_{i}
\end{equation}
are as follows:

\begin{equation}
\alpha_{i} = \frac{1}{\sqrt{2\pi}\left[1+\textrm{erf}\left(\frac{\mu_{d,i}}{\sqrt{2\left(\sigma_{d,i}^{2} + 1\right)}}\right)\right]}
\end{equation}

\begin{equation} 
\mu_{i} = -\frac{2\alpha_{i}}{\sqrt{\sigma_{d,i}^2 + 1}}\exp\left(-\frac{\mu_{d,i}^{2}}{2(\sigma_{d,i}^2 + 1)}\right)
\end{equation}

\begin{equation} 
\begin{split}
\sigma_{i}^{2} &= \alpha_{i}\left[\sqrt{2\pi}\left(\left(1+\mu_{i}^{2}\right)\left(1+\textrm{erf}\left(\frac{\mu_{d,i}}{\sqrt{2(\sigma_{d,i}^{2}+1)}}\right)\right)\right) \right. \\
& \qquad - \left. \frac{2}{\sqrt{\sigma_{d,i}^2 + 1}}\exp\left(-\frac{\mu_{d,i}^{2}}{2(\sigma_{d,i}^2 + 1)}\right)\left(\frac{\mu_{d,i}}{\sigma_{d,i}^2 + 1} - 2\mu_{i} \right)\right]
\end{split}
\end{equation}

\begin{figure}
  \centering
  \subfloat[$\mu_{c,i} = -2, \sigma_{c,i} = 0.5$]{
    \includegraphics[width=0.45\textwidth]{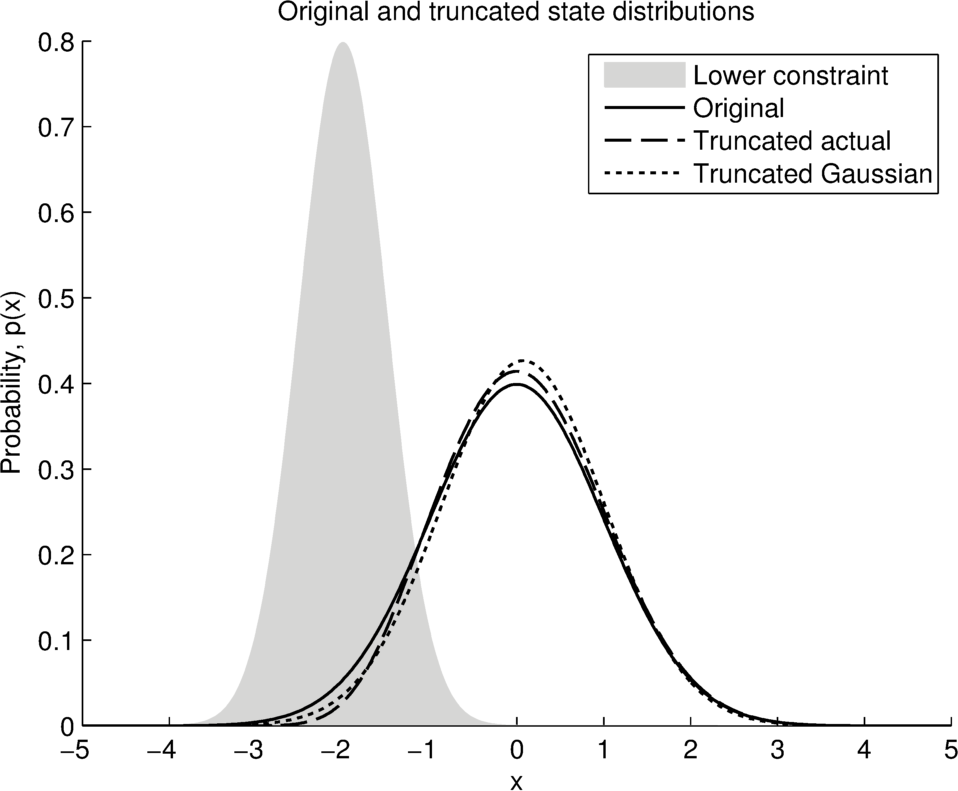}\label{sf:single1}
  }
  \hspace{0.5em}
  \subfloat[$\mu_{c,i} = 0, \sigma_{c,i} = 1$]{
    \includegraphics[width=0.45\textwidth]{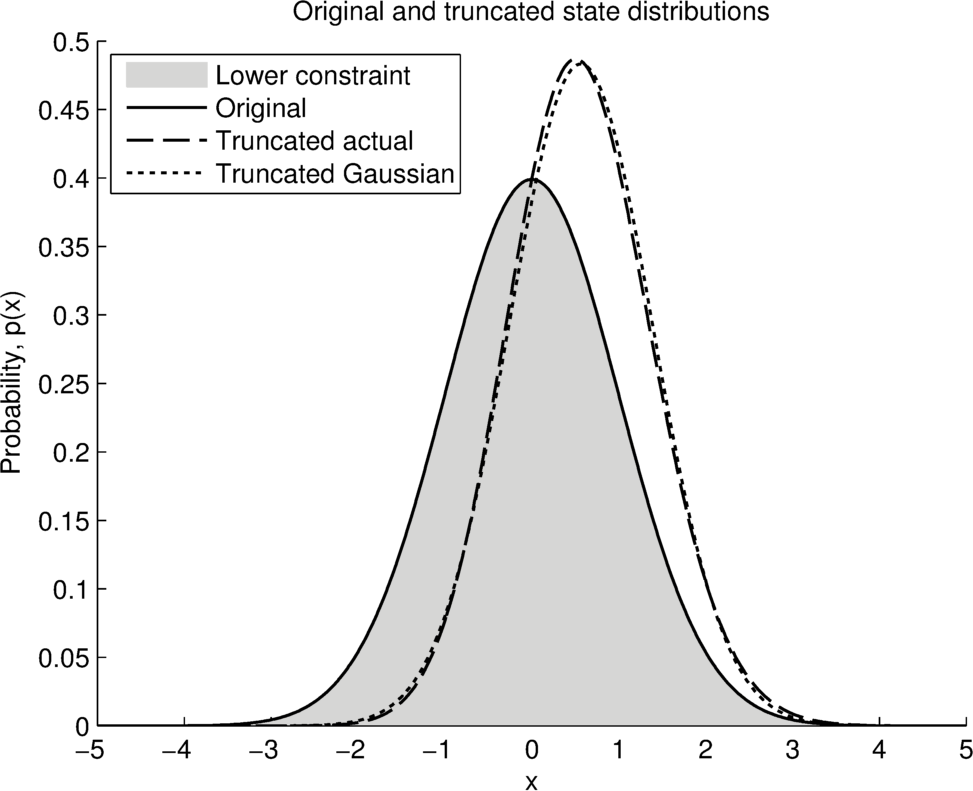}\label{sf:single2}
  }
  
  \subfloat[$\mu_{c,i} = 3, \sigma_{c,i} = 1.5$]{
    \includegraphics[width=0.45\textwidth]{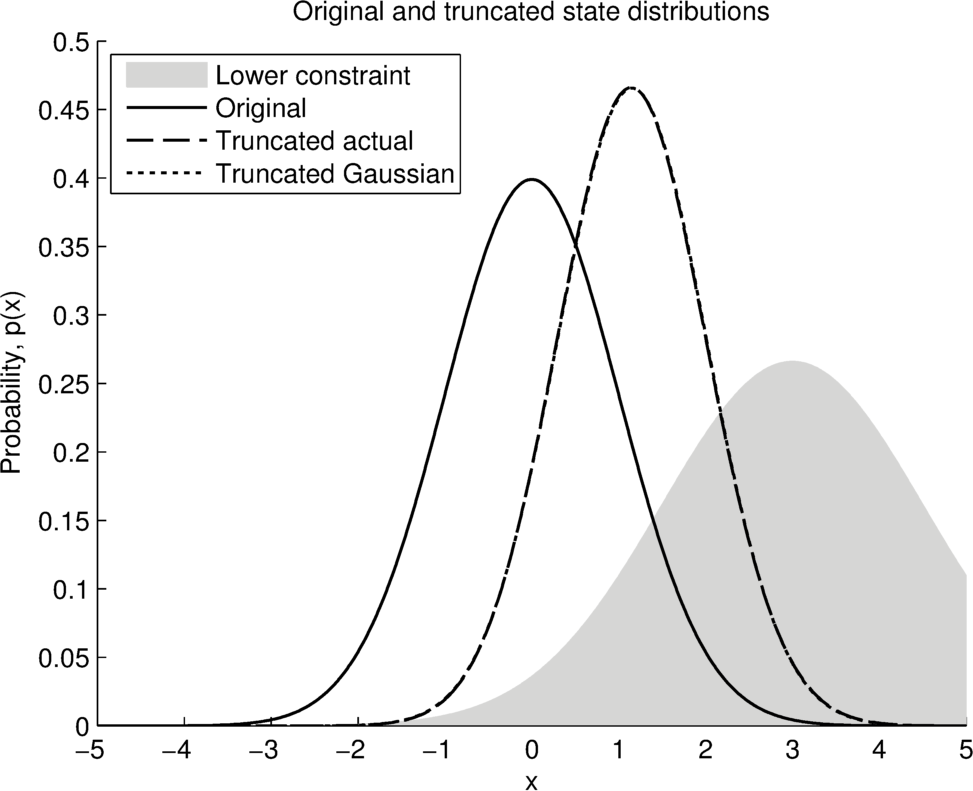}\label{sf:single3}
  }
  \caption{Comparison of the actual truncated distributions and Gaussian approximations of the truncated distributions for several lower constraints. In \protect\subref{sf:single3}, the lower constraint is higher than the original distribution, resulting in the truncated distribution moving towards the constraint. In this case, the Gaussian approximation is an almost perfect approximation of the truncated distribution and the two lines overlap. }
  \label{f:single_examples}
\end{figure}

Several examples of the proposed method with a lower constraint are shown in Figure \ref{f:single_examples}. As can be seen, the Gaussian approximation is very close to the actual truncated distribution, with the approximation improving as $\sigma_{c,i}$ increases. As $\sigma_{c,i} \rightarrow \infty$, the $\textrm{CDF}\left(C_{i}\right)$ approaches a uniform distribution, which means that the truncated PDF approaches the original PDF. In Figure \ref{sf:single3}, the lower constraint is higher than the original distribution, resulting in the truncated distribution moving towards the constraint. In this case, the truncated distribution is actually still below the majority of the constraint distribution. Here, the uncertainty of the original and constraint distributions are balanced against one another---the more certain that one of the distributions is, the closer the truncated distribution will be to that distribution. For example, as the uncertainty of the constraint is decreased, the truncated distribution will move to the right. As the uncertainty of the constraint approaches 0, the constraint approaches a hard constraint and the majority of the truncated distribution will be above the constraint. 

\subsection{Feedback of the truncated estimate}

\begin{figure}
  \centering
  \subfloat[The truncated state estimate is fed back into the Kalman filter.]{
    \includegraphics[width=0.8\textwidth]{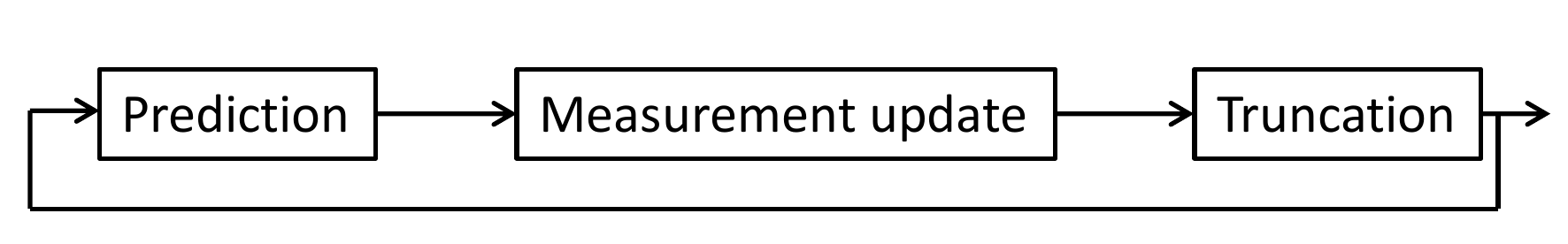}\label{sf:feedback_flowchart}
  }
  
  \subfloat[The Kalman filter is run independently of the truncation, with the feedback occuring after the measurement update and before the truncation. ]{
    \includegraphics[width=0.8\textwidth]{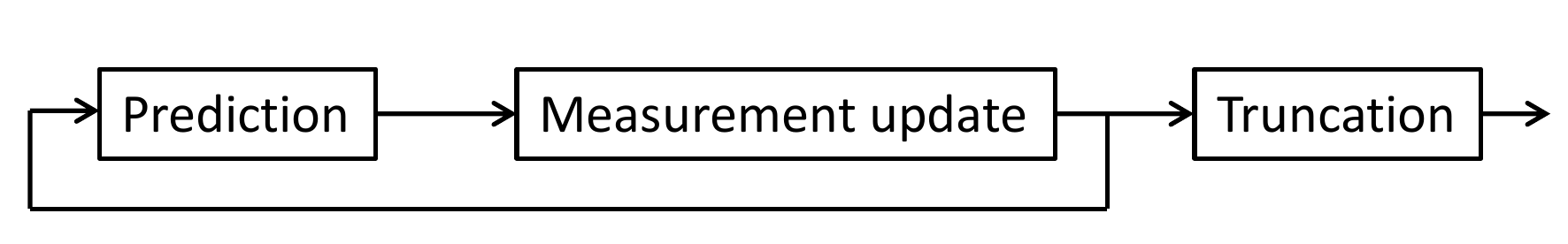}\label{sf:no_feedback_flowchart}
  }
  \caption{Feedback of the state estimate into the Kalman filter can occur either before or after the truncation process. Deciding which feedback approach to use depends on the uncertainty model of the constraints and the shape of the truncated distribution. }
  \label{f:feedback_flowcharts}
\end{figure}

There is disagreement amongst authors as to whether or not the truncated state estimate should be fed back into the Kalman filter, with some suggesting using feedback \cite{Teixeira2010, Straka2012} as shown in Figure~\ref{sf:feedback_flowchart} and others stating that the truncation process should be kept independent of the unconstrained Kalman filter \cite{Simon2010b} as shown in Figure~\ref{sf:no_feedback_flowchart}. In \cite{Simon2010b}, it is argued that this feedback can lead to overconfident estimates as the information provided by the constraints is used multiple times. In reality, there are two issues to consider when deciding whether or not to use feedback. The first issue concerns the uncertainty model of the constraints---if the constraints are noisy, and if that noise is independent from one time-step to another, then the truncated estimate can be fed back into the Kalman filter. Under these conditions, the constraints are similar to independent noisy measurements. However, many physical constraints are uncertain rather than noisy---that is, the actual value of the constraint is constant and is not resampled at each time-step. Feeding the truncated estimate back into the Kalman filter in this case can result in overconfident estimates, as will be shown in Section \ref{s:results}. 

The second issue arises when the truncated distribution is highly non-Gaussian, which is commonly the case when the uncertainty of the constraint is low in comparison to the uncertainty of the estimate. In these cases, the Gaussian approximation of the truncated distribution introduces error that can accumulate if it is fed back into the Kalman filter, leading to unrepresentative state estimates. An example of such a situation is given in \cite{Simon2010b}. Consider a scalar system with no process noise such that $x_{k+1} = x_{k}$, no measurements, and a hard constraint of $x \ge 0$. If the initial state estimate is a standard normal distribution, then the truncated distribution will have a Gaussian shape for $x \ge 0$ and 0 otherwise. Approximating this as a Gaussian distribution changes the mean from 0 to $\sqrt{2/\pi}$ and the variance from 1 to $\left(\pi-2\right)/\pi$ after the truncation has been applied once. If this were fed back into the Kalman filter, it would result in a monotonically increasing estimate mean and a monotonically decreasing estimate variance for successive applications of the truncation approach. The authors of \cite{Simon2010b} argue that this is a result of incorporating the information from the constraints multiple times. However, it is actually the Gaussian approximation of the truncated distribution that causes this behaviour. If it were possible to feed the truncated distribution back into the filter without approximating it as a Gaussian distribution, then applying the constraint at the next time step would result in the exact same truncated distribution. 

When deciding whether or not to use feedback, the above issues should be carefully considered. Not using feedback is the conservative option---the resultant estimates will have a higher uncertainty compared to methods using feedback, but will avoid most of these issues. Provided the constraints are noisy rather than uncertain, and have a high noise uncertainty, using feedback is valid and can yield a significant performance benefit over not using feedback. The difference between uncertain and noisy constraints is minimal when the uncertainty of the constraints is small in comparison to the uncertainty of the state estimate, and the main source of error in these cases is the approximation of the truncated distribution. One possible way of dealing with this, suggested by \cite{Simon2010b}, is to only feed back the elements of the truncated estimate where the elements of the original estimate violate the constraint. This was suggested in the context of hard constraints, however, and determining the point at which a soft constraint is violated is ambiguous and is left to the reader. 

\section{Interval constraint}\label{s:interval}

Now consider the case where there are two constraints. After transforming the state using Equations (\ref{eq:transform_start})--(\ref{eq:transform_end}), the two constraints acting on the transformed state are:

\begin{equation}
C_{i} \le \boldsymbol{e}_{i}^{T}\boldsymbol{z}_{i} \le D_{i}
\end{equation}

Using Bayes' theorem, the conditional probability distribution of the transformed state satisfying the constraints, $p_{\boldsymbol{e}_{i}^{T}\boldsymbol{z}_{i}}(\zeta | (C_{i} \le \zeta \le D_{i}))$, as a function of $\zeta$ is given by:

\begin{align} 
p_{\boldsymbol{e}_{i}^{T}\boldsymbol{z}_{i}}(\zeta | C_{i} \le \zeta &\le D_{i}) = \frac{p_{\boldsymbol{e}_{i}^{T}\boldsymbol{z}_{i}}(\zeta) \times P(C_{i} \le \zeta \le D_{i})} {P(C_{i} \le \boldsymbol{e}_{i}^{T}\boldsymbol{z}_{i} \le D_{i})} \label{eq:bayes_interval1} \\ 
& = \frac{p_{\boldsymbol{e}_{i}^{T}\boldsymbol{z}_{i}}(\zeta) \times P(C_{i} \le \zeta) \times P( \zeta \le D_{i})} {P(C_{i} \le \boldsymbol{e}_{i}^{T}\boldsymbol{z}_{i} \le D_{i})} \label{eq:bayes_interval2}
\end{align}

Replacing $P(C_{i} \le \zeta \le D_{i})$ in (\ref{eq:bayes_interval1}) with $P(C_{i} \le \zeta) \times P( \zeta \le D_{i})$  in (\ref{eq:bayes_interval2}) is possible since $C_{i}$ and $D_{i}$ are independent random variables. Note, however, that $P(C_{i} \le \boldsymbol{e}_{i}^{T}\boldsymbol{z}_{i} \le D_{i})$ cannot be split up in this way and requires the evaluation of a multi-variate CDF, which does not have an analytical solution. This probability is a normalising factor for the numerator. 

This gives the following distribution:

\begin{align} 
p_{\boldsymbol{e}_{i}^{T}\boldsymbol{z}_{i}}(\zeta &| C_{i} \le \zeta \le D_{i}) \notag \\
& = \frac{p_{\boldsymbol{e}_{i}^{T}\boldsymbol{z}_{i}}(\zeta) \times P(C_{i} \le \zeta) \times P( \zeta \le D_{i})} {P(C_{i} \le \boldsymbol{e}_{i}^{T}\boldsymbol{z}_{i} \le D_{i})} \notag \\
& = \frac{\textrm{PDF}_{\boldsymbol{e}_{i}^{T}\boldsymbol{z}_{i}}(\zeta) \times \textrm{CDF}_{C_{i}}(\zeta)  \times \textrm{CDF}_{D_{i}}(\zeta)}{P(C_{i} \le \boldsymbol{e}_{i}^{T}\boldsymbol{z}_{i} \le D_{i})} \notag \\
& = \frac{\frac{1}{\sqrt{2\pi}}\exp(-\frac{\zeta^{2}}{2})\frac{1}{2}\left[ 1 + \textrm{erf}\left( \frac{\zeta - \mu_{c,i}}{\sigma_{c,i}\sqrt{2}} \right) \right]\frac{1}{2}\left[ 1 - \textrm{erf}\left( \frac{\zeta - \mu_{d,i}}{\sigma_{d,i}\sqrt{2}} \right) \right]}{P(C_{i} \le \boldsymbol{e}_{i}^{T}\boldsymbol{z}_{i} \le D_{i})} \label{eq:dist_int}
\end{align}


The integrals required to calculate the mean and variance of (\ref{eq:dist_int}) contain integrals of the form:

\begin{equation}
\int\limits_{-\infty}^{\infty}\exp\left(-x^2\right)\textrm{erf}(a x + b)\textrm{erf}(c x + d) \textrm{d}x
\end{equation}
which does not have an analytical solution. The following approximation is proposed:

\begin{align}
&\left[ 1 + \textrm{erf}\left( \frac{\zeta - \mu_{c,i}}{\sigma_{c,i}\sqrt{2}} \right) \right]\left[ 1 - \textrm{erf}\left( \frac{\zeta - \mu_{d,i}}{\sigma_{d,i}\sqrt{2}} \right) \right] \notag \\
& \approx 2 \left[\textrm{erf}\left( \frac{\zeta - \mu_{c,i}}{\sigma_{c,i}\sqrt{2}} \right) -  \textrm{erf}\left( \frac{\zeta - \mu_{d,i}}{\sigma_{d,i}\sqrt{2}} \right)  \right] \label{eq:approximation}
\end{align}

\begin{figure}
\centering
\includegraphics[width = 0.6\textwidth]{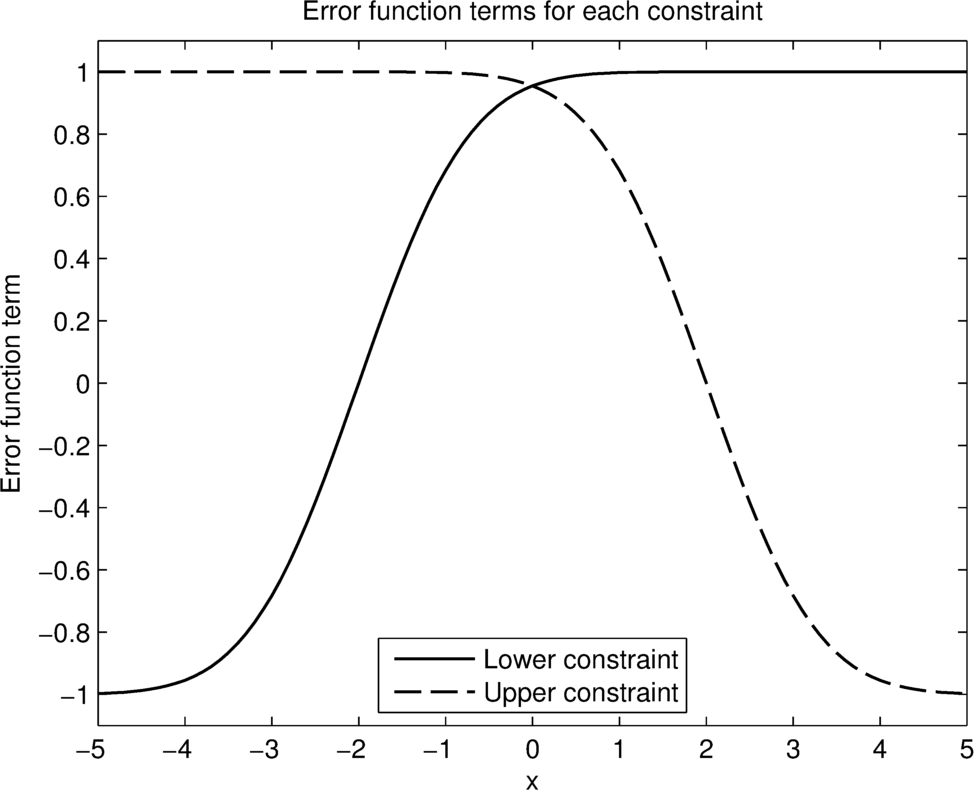}
\caption{Error function terms for a lower constraint with $\mu_{c,i} = -2$ and $\sigma_{c,i} = 1$ and an upper constraint with $\mu_{d,i} = 2$ and $\sigma_{d,i} = 1$. In this case, one of the error function terms is always close to 1. } \label{f:erf}
\end{figure}

This yields an approximation of the numerator in (\ref{eq:dist_int}) of:

\begin{equation}\label{eq:approximate_pdf}
\textrm{Z}(\boldsymbol{e}_{i}^{T}\boldsymbol{\tilde{z}}_{i+1}) = \frac{1}{4\sqrt{2\pi}}\exp\left(-\zeta^{2}/2\right)\left[2\left(\textrm{erf}\left(\frac{\zeta-\mu_{c,i}}{\sigma_{c,i}\sqrt{2}} \right)-\textrm{erf}\left(\frac{\zeta-\mu_{d,i}}{\sigma_{d,i}\sqrt{2}} \right)\right)\right]
\end{equation}
where $\textrm{Z}(\boldsymbol{e}_{i}^{T}\boldsymbol{\tilde{z}}_{i+1})$ is an unnormalised function describing the truncated distribution. This approximation relies on the condition that $\mu_{c,i} < \mu_{d,i}$, and the assumption that the constraint distributions do not significantly overlap. If these are satisfied, it is highly likely that one of the erf terms will be equal to 1 when the other is not, giving a good approximation of the actual distribution. This is illustrated in Figure \ref{f:erf}. 

An overlap metric, $\gamma$, is defined as:

\begin{equation} \label{eq:gamma}
\gamma = \frac{\mu_{d,i} - \mu_{c,i}}{\sigma_{d,i} + \sigma_{c,i}}
\end{equation}
and a shape metric, $\delta$, is defined as:

\begin{equation} \label{eq:delta}
\delta = \left| \log\left(\frac{\sigma_{c,i}}{\sigma_{d,i}}\right)\right|
\end{equation}

\begin{figure}
  \centering
  \subfloat[$\mu_{c,i} = -2, \sigma_{c,i} = 0.5, \mu_{d,i} = 2, \sigma_{d,i} = 1$, corresponding to $\gamma = 2.67$ and $\delta = 0.30$]{
    \includegraphics[width=0.45\textwidth]{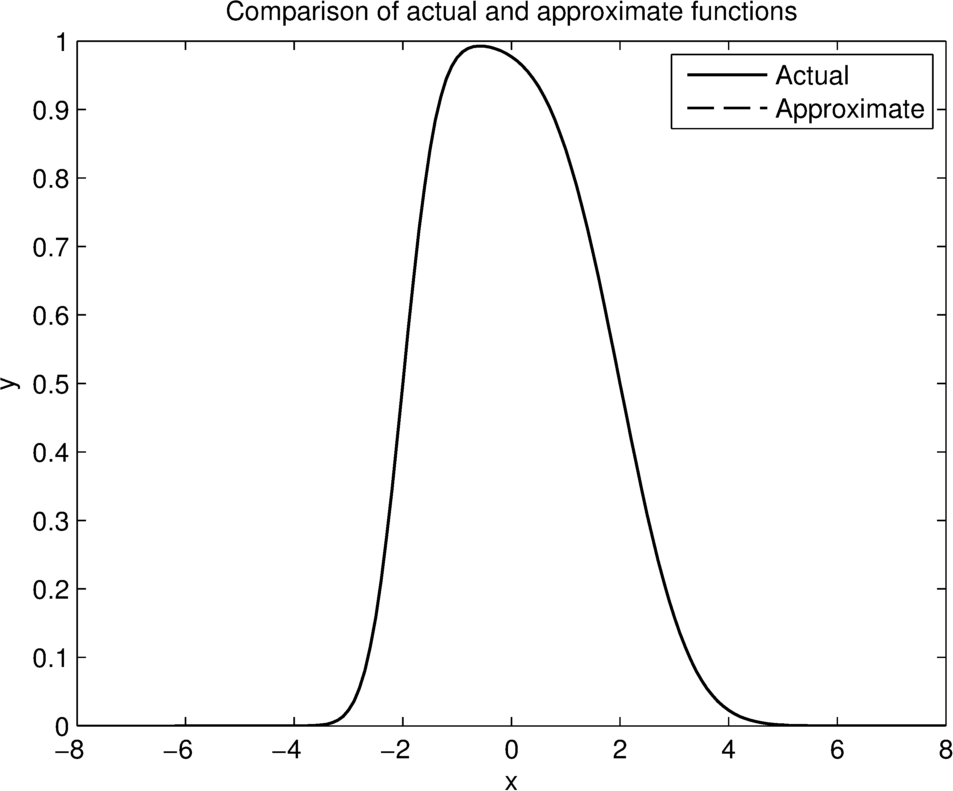}\label{sf:approx1}
  }
  \hspace{0.5em}
  \subfloat[$\mu_{c,i} = -3, \sigma_{c,i} = 1, \mu_{d,i} = 2, \sigma_{d,i} = 3$, corresponding to $\gamma = 1.25$ and $\delta = 0.48$]{
    \includegraphics[width=0.45\textwidth]{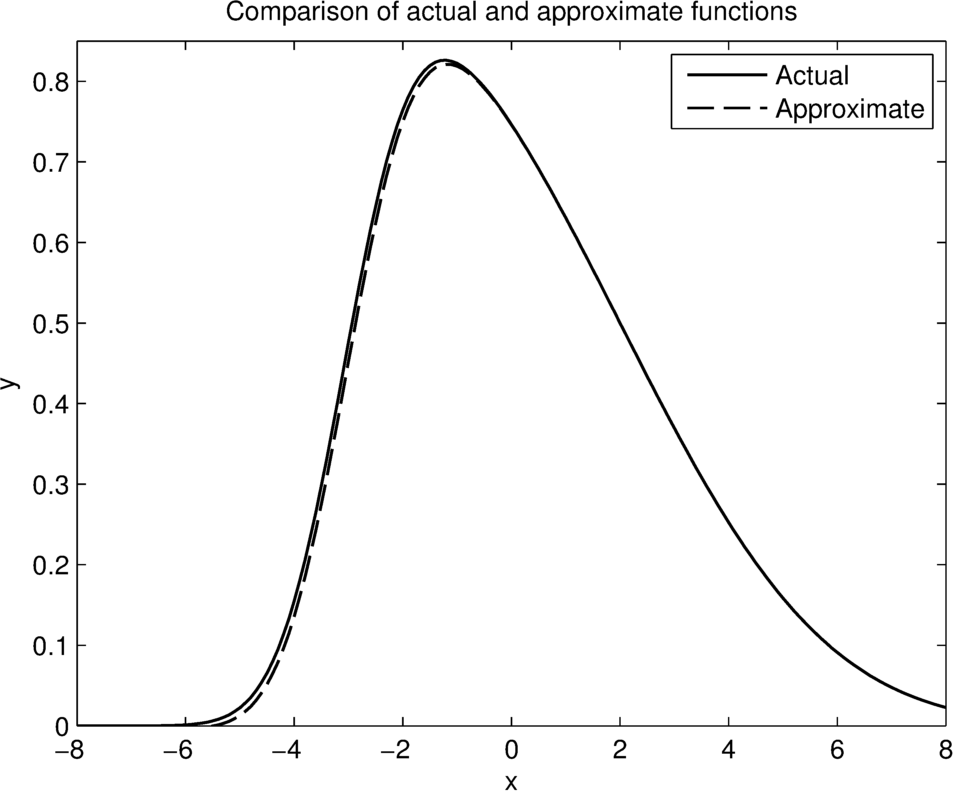}\label{sf:approx2}
  }
  
  \subfloat[$\mu_{c,i} = -1, \sigma_{c,i} = 2, \mu_{d,i} = 2, \sigma_{d,i} = 3.5$, corresponding to $\gamma = 0.55$ and $\delta = 0.24$]{
    \includegraphics[width=0.45\textwidth]{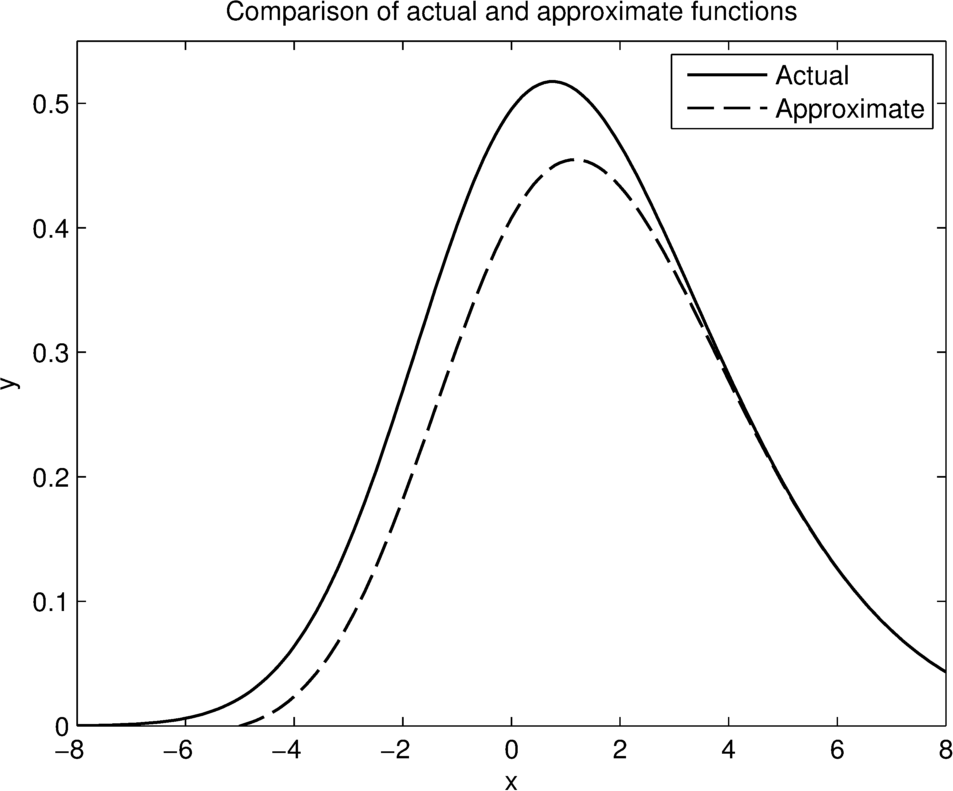}\label{sf:approx3}
  }
  \caption{Comparison of the actual and approximate functions from (\ref{eq:approximation}) for several values of $\gamma$ and $\delta$. In \protect\subref{sf:approx1}, the approximation is an almost perfect approximation of the actual distribution and the two lines overlap. }
  \label{f:approx_examples}
\end{figure}

$\gamma$ is a measure of how much the probability distributions of the two constraints overlap, and $\delta$ is a measure of how different the shapes of the probability distributions of the constraints are. Figure \ref{f:approx_examples} shows examples of the approximation applied to several different cases of $\gamma$ and $\delta$. As can be seen, the approximation is an almost perfect approximation in Figure \ref{sf:approx1}, with the approximation degrading as $\gamma$ decreases in the other examples. 

\begin{figure}
  \centering
    \includegraphics[width=0.6\textwidth]{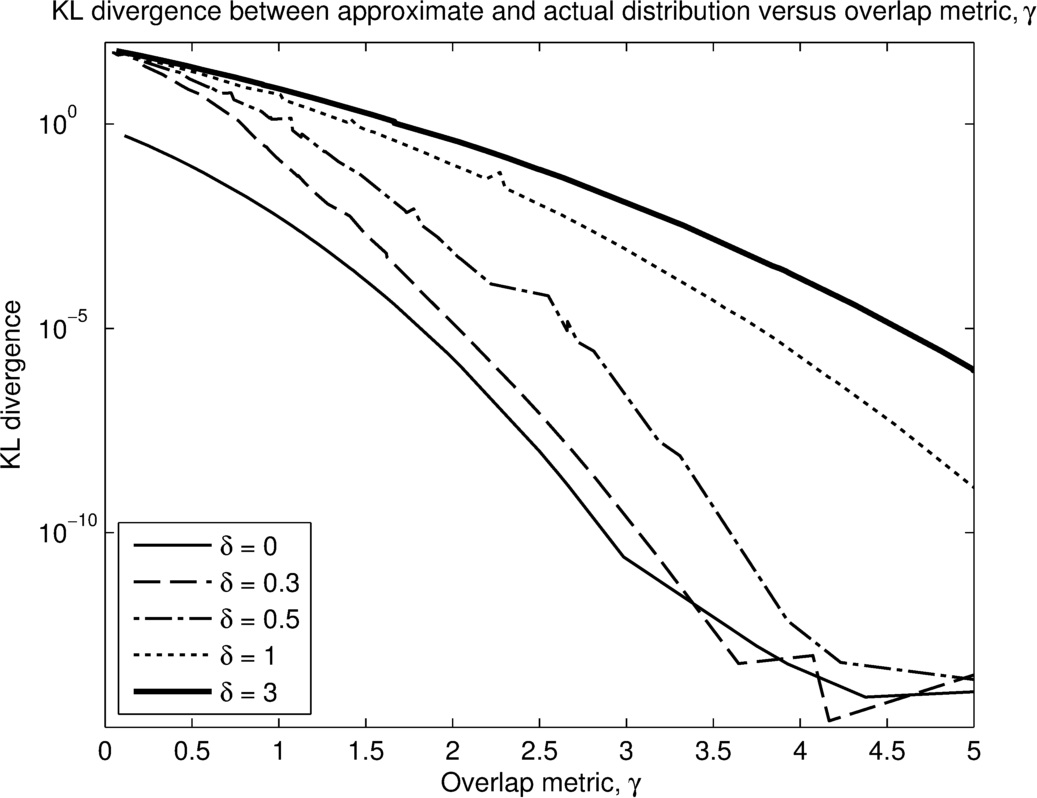}
  \caption{Comparison of the KL divergence between the actual and approximate probability distributions as a function of the overlap metric, $\gamma$, where the lines are constant $\delta$. For $\delta > 3$, the KL divergence is approximately the same as for $\delta = 3$. }
  \label{f:approx_error}
\end{figure}

Figure \ref{f:approx_error} shows the Kullback--Leibler (KL) divergence between the actual and approximate probability distributions for various $\delta$ and $\gamma$. Increasing $\gamma$ and decreasing $\delta$ improves the approximation. For $\gamma \ge 3$, the approximation very closely matches the actual distribution, regardless of $\delta$. 

To normalise $\textrm{Z}(\boldsymbol{e}_{i}^{T}\boldsymbol{\tilde{z}}_{i+1})$ to a PDF, the area of the function is calculated as:

\begin{equation}
\begin{aligned}
\int\limits_{-\infty}^{\infty}\textrm{Z}&(\boldsymbol{e}_{i}^{T}\boldsymbol{\tilde{z}}_{i+1}) \textrm{d}\zeta \\
&= \int\limits_{-\infty}^{\infty}\frac{1}{4\sqrt{2\pi}}\exp\left(-\zeta^{2}/2\right)\left[2\left(\textrm{erf}\left(\frac{\zeta-\mu_{c,i}}{\sigma_{c,i}\sqrt{2}} \right)-\textrm{erf}\left(\frac{\zeta-\mu_{d,i}}{\sigma_{d,i}\sqrt{2}} \right)\right)\right] \textrm{d}\zeta \\
&= \frac{1}{2}\left[\textrm{erf}\left(\frac{\mu_{d,i}}{\sqrt{2\left(\sigma_{d,i}^{2} + 1\right)}}\right) - \textrm{erf}\left(\frac{\mu_{c,i}}{\sqrt{2\left(\sigma_{c,i}^{2} + 1\right)}}\right)\right]
\end{aligned}
\end{equation}

The mean of the PDF is then given by:

\begin{align}
\mu_{i} &= \alpha_{i}\int\limits_{-\infty}^{\infty}\zeta \exp\left(-\zeta^{2}/2\right)\left[\left(\textrm{erf}\left(\frac{\zeta-\mu_{c,i}}{\sigma_{c,i}\sqrt{2}} \right)-\textrm{erf}\left(\frac{\zeta-\mu_{d,i}}{\sigma_{d,i}\sqrt{2}} \right)\right)\right] \textrm{d}\zeta    \notag\\
&= 2\alpha_{i}\left(\frac{1}{\sqrt{\sigma_{c,i}^2 + 1}}\exp\left(-\frac{\mu_{c,i}^{2}}{2(\sigma_{c,i}^2 + 1)}\right) \right. \notag \\
&\qquad \left. - \frac{1}{\sqrt{\sigma_{d,i}^2 + 1}}\exp\left(-\frac{\mu_{d,i}^{2}}{2(\sigma_{d,i}^2 + 1)}\right)\right)
\end{align}
where
\begin{equation}
\alpha_{i} = \frac{1}{\sqrt{2\pi}\left[\textrm{erf}\left(\frac{\mu_{d,i}}{\sqrt{2\left(\sigma_{d,i}^{2} + 1\right)}}\right) - \textrm{erf}\left(\frac{\mu_{c,i}}{\sqrt{2\left(\sigma_{c,i}^{2} + 1\right)}}\right)\right]}
\end{equation}

The variance is given by:

\begin{equation}
\begin{aligned}
\sigma_{i}^{2} = \alpha_{i}\int\limits_{-\infty}^{\infty}\left(\zeta - \mu_{i}\right)^{2} \exp\left(-\zeta^{2}/2\right)\left[\left(\textrm{erf}\left(\frac{\zeta-\mu_{c,i}}{\sigma_{c,i}\sqrt{2}} \right)-\textrm{erf}\left(\frac{\zeta-\mu_{d,i}}{\sigma_{d,i}\sqrt{2}} \right)\right)\right] \textrm{d}\zeta    \\
=  \alpha_{i}\left[\sqrt{2\pi}\left(\left(1+\mu_{i}^{2}\right)\left(\textrm{erf}\left(\frac{\mu_{d,i}}{\sqrt{2(\sigma_{d,i}^{2}+1)}}\right)-\textrm{erf}\left(\frac{\mu_{c,i}}{\sqrt{2(\sigma_{c,i}^{2}+1)}}\right)\right)\right) \right. \\ 
 \qquad + \left. \frac{2}{\sqrt{\sigma_{c,i}^2 + 1}}\exp\left(-\frac{\mu_{c,i}^{2}}{2(\sigma_{c,i}^2 + 1)}\right)\left(\frac{\mu_{c,i}}{\sigma_{c,i}^2 + 1} - 2\mu_{i} \right) \right. \\
 \qquad - \left. \frac{2}{\sqrt{\sigma_{d,i}^2 + 1}}\exp\left({-\frac{\mu_{d,i}^{2}}{2(\sigma_{d,i}^2 + 1)}}\right)\left(\frac{\mu_{d,i}}{\sigma_{d,i}^2 + 1} - 2\mu_{i} \right)\right]
\end{aligned}
\end{equation}

The derivations of these integrals are not provided, but can be easily derived using the solutions provided in \ref{s:a_mean} and \ref{s:a_variance} for the one-sided constraint. The truncated state estimate is then obtained by applying Equations (\ref{eq:mean_and_sigma}) and (\ref{eq:inverse_transform}). This process is repeated for the $s$ constraints, incrementing $i$ each time. 

\begin{figure}
  \centering
  \subfloat[$\mu_{c,i} = -2, \sigma_{c,i} = 0.5, \mu_{d,i} = 2, \sigma_{d,i} = 1$]{
    \includegraphics[width=0.45\textwidth]{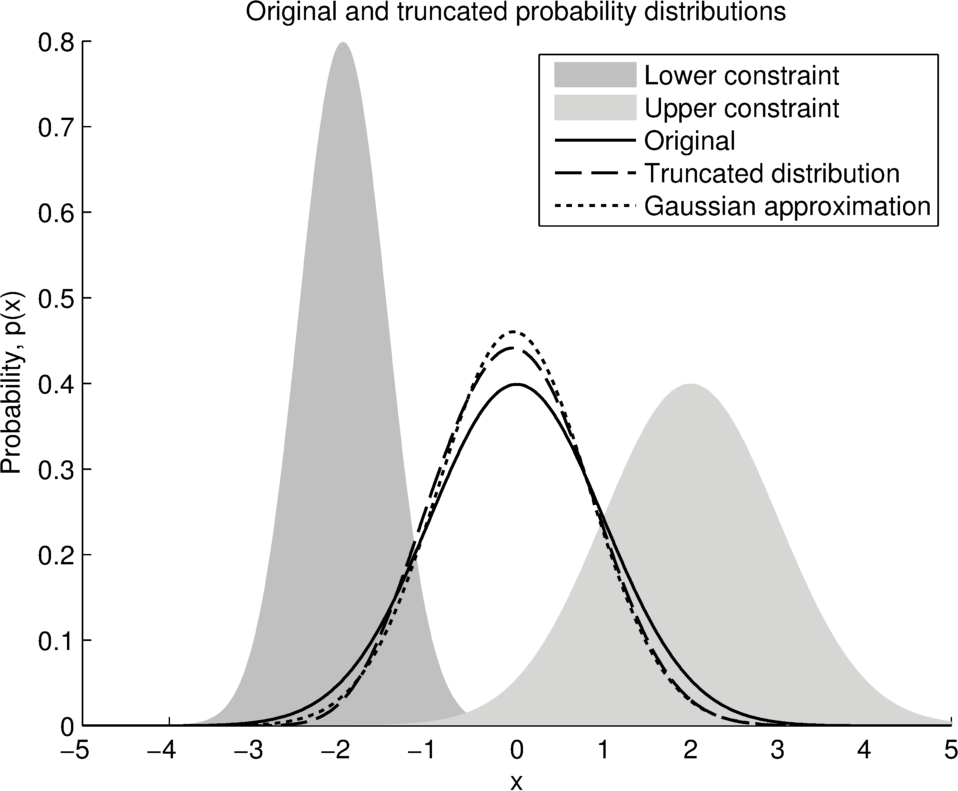}\label{sf:interval1}
  }
  \hspace{0.5em}
  \subfloat[$\mu_{c,i} = 1, \sigma_{c,i} = 1, \mu_{d,i} = 3, \sigma_{d,i} = 1$]{
    \includegraphics[width=0.45\textwidth]{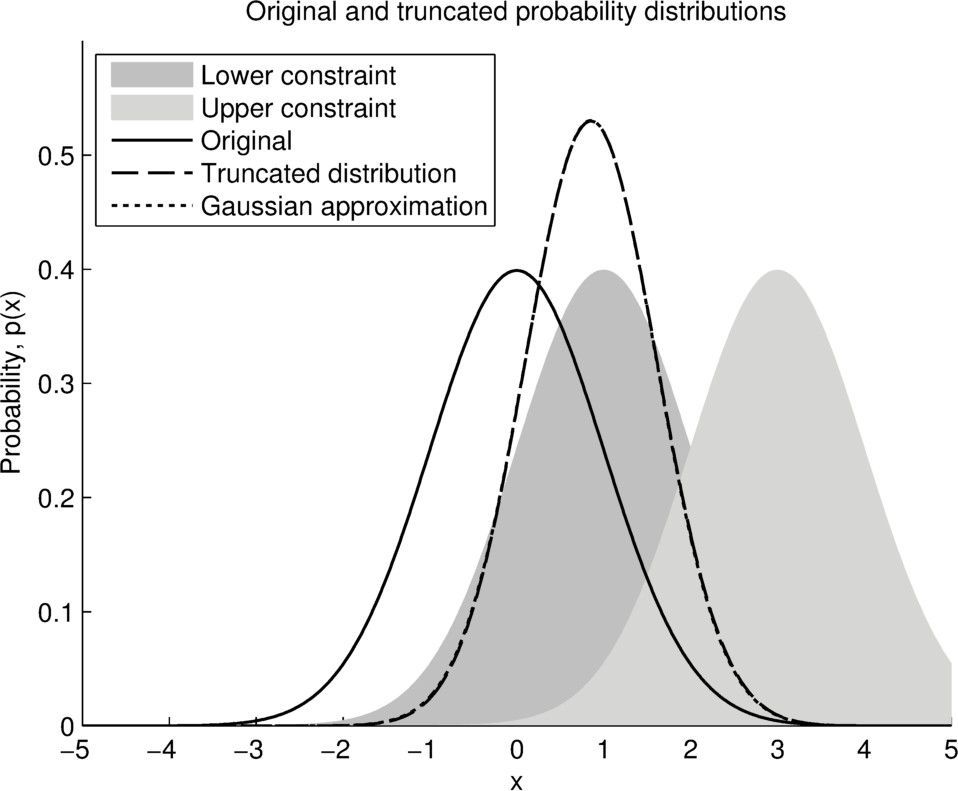}\label{sf:interval2}
  }
  
  \subfloat[$\mu_{c,i}=-2, \sigma_{c,i}=0.001, \mu_{d,i}=-1, \sigma_{d,i}=0.5$]{
    \includegraphics[width=0.45\textwidth]{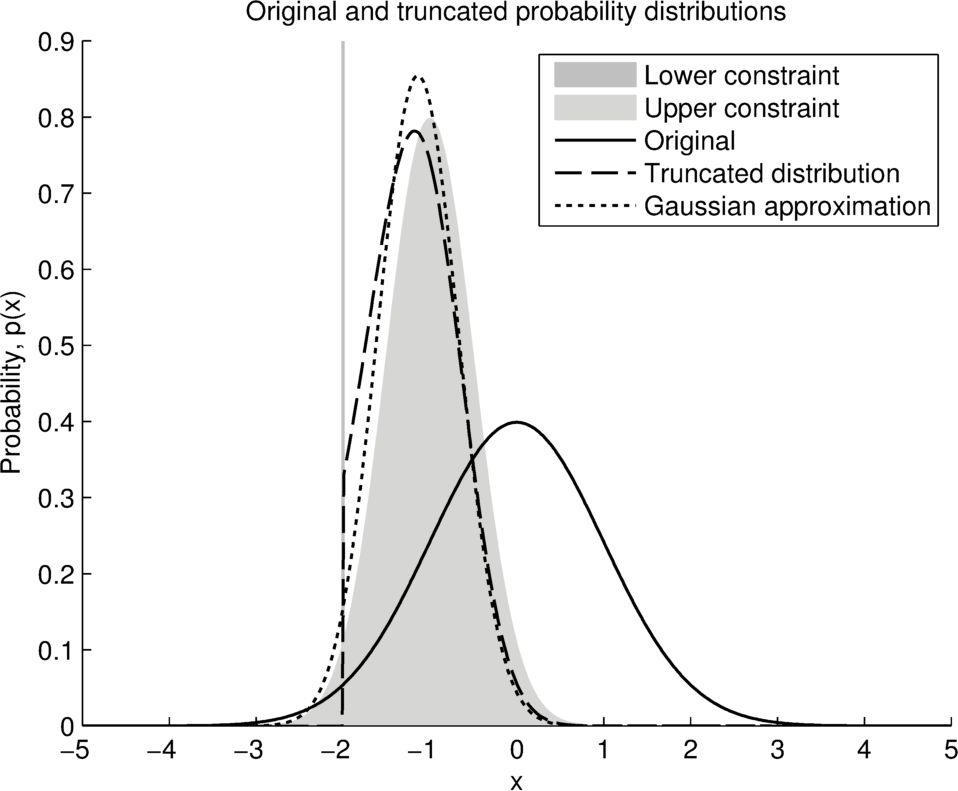}\label{sf:interval3}
  }
  \caption{Comparison of the actual and approximate distributions for several values of $\gamma$ and $\delta$. In \protect\subref{sf:interval2}, the Gaussian approximation is an almost perfect approximation of the truncated distribution and the two lines overlap. }
  \label{f:interval_examples}
\end{figure}

Several examples of the proposed method for interval constraints are shown in Figure \ref{f:interval_examples}. The Gaussian approximation method produces distributions that are very similar to the actual truncated distributions. Figure \ref{sf:interval3} shows an example where soft and hard constraints are combined---the hard constraint has been modelled as a soft constraint with a very small standard deviation. 

\section{Results}\label{s:results}

Consider a robot moving along a corridor, as shown in Figure \ref{f:robotexample}. The corridor has a wall 10m in front of the initial position of the robot, and discrete position sensors placed at 1m intervals. The position sensors can detect which side of the set-point of the sensor the robot is on, and have uncertainty on the set-point. The robot has an initial velocity of 10cm/s and accelerates at 1cm/s$^{2}$ for 20s, then decelerates at 1cm/s$^{2}$ for 20s, before again accelerating at 1cm/s$^{2}$ until it reaches the far wall. Two types of robots were considered---one with standard deviations on the acceleration and initial velocity of $\sigma_{a} = 1\textrm{cm/s}^{2}$ and $\sigma_{v} = 3 \textrm{cm/s}$ respectively (Robot A), and a less uncertain one with standard deviations on the acceleration and initial velocity of $\sigma_{a} = 0.5\textrm{cm/s}^{2}$ and $\sigma_{v} = 1.5 \textrm{cm/s}$ respectively (Robot B). The standard deviation of the set-point of the sensors was also varied, with standard deviations (in cm) of $\sigma_{s} \in \{0, 5, 10, 15, 20, 25, 30\}$ tested. The position of each robot was tracked using the following Kalman filter run at 10Hz:

\begin{figure}
\centering
\includegraphics[width = 0.5\textwidth]{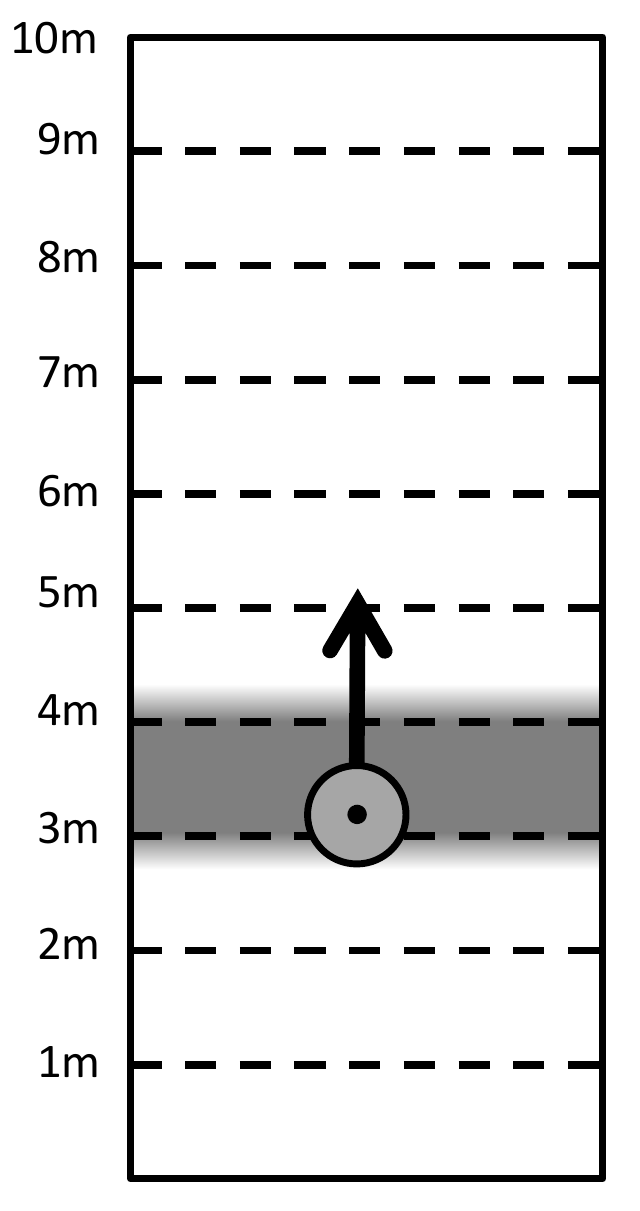}
\caption{The robot (circle) is moving along the corridor. Positioned at 1m intervals are sensors with uncertain positions that can detect which side of the sensor the robot is on. These sensor readings are used as both measurements (when the sensor reading changes) and constraints. In the image, the position estimate of the robot would be constrained between the sensors at 3m and 4m (shaded area). The uncertainty of the constraints is shown by the shading. } \label{f:robotexample}
\end{figure}

\begin{equation}
\boldsymbol{x} = \begin{bmatrix}
x \\
\dot{x}
\end{bmatrix} \quad \boldsymbol{u} = \begin{bmatrix}
\ddot{x}
\end{bmatrix}\quad \boldsymbol{F} = \begin{bmatrix}
1 & \Delta t\\
0 & 1 
\end{bmatrix} \quad \boldsymbol{G} = \begin{bmatrix}
\frac{\Delta t^{2}}{2}\\
\Delta t
\end{bmatrix}
\end{equation}

The covariance of the process noise was given by:

\begin{equation}
\boldsymbol{Q} = \boldsymbol{G}\boldsymbol{G}^{T}\sigma_{a}^{2} = \begin{bmatrix}
\frac{\Delta t^{4}}{4} & \frac{\Delta t^{3}}{2}\\
\frac{\Delta t^{3}}{2} & \Delta t^{2}
\end{bmatrix} \sigma_{a}^{2}
\end{equation}

The Kalman filter was initialised with:

\begin{equation}
\hat{\boldsymbol{x}} = \begin{bmatrix}
0\\
0.1
\end{bmatrix} \quad \boldsymbol{P} = \begin{bmatrix}
0 & 0 \\
0 & \sigma_{v}^{2}
\end{bmatrix}
\end{equation}

A position sensor changing its reading was incorporated as a noisy measurement of the position:

\begin{equation}
\boldsymbol{H} = \begin{bmatrix}
1 & 0
\end{bmatrix} \quad \boldsymbol{R} = \begin{bmatrix}
\sigma_{s}^{2}
\end{bmatrix}
\end{equation}

At each time-step, position sensors whose reading did not change were not incorporated into the Kalman filter. The aim of the constrained Kalman filtering approach here was to use the absence of measurements to improve the state estimate. While the robot was in between sensors, the sensors were treated as constraints on the state of the system. The truncation method proposed in this paper (which will be referred to as the soft-constrained Kalman filter) was compared with an unconstrained Kalman filter, and a constrained Kalman filter using the truncation method that ignored the uncertainty of the constraints and treated them as hard constraints (referred to as the hard-constrained Kalman filter). Each combination of robot, sensor uncertainty, and Kalman filter method was tested 1000 times. 

\begin{figure}
  \centering
  \subfloat[RMSE]{
    \includegraphics[width=0.6\textwidth]{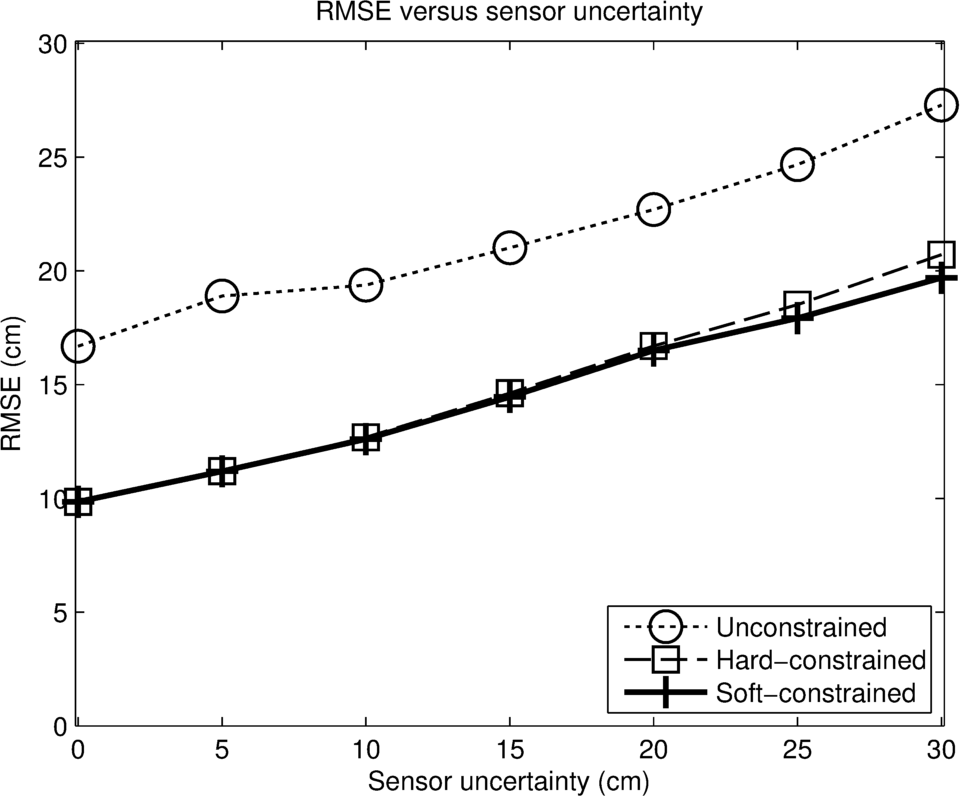}\label{sf:results1_rmse}
  }
  
  \subfloat[Percentage improvement between methods]{
    \includegraphics[width=0.6\textwidth]{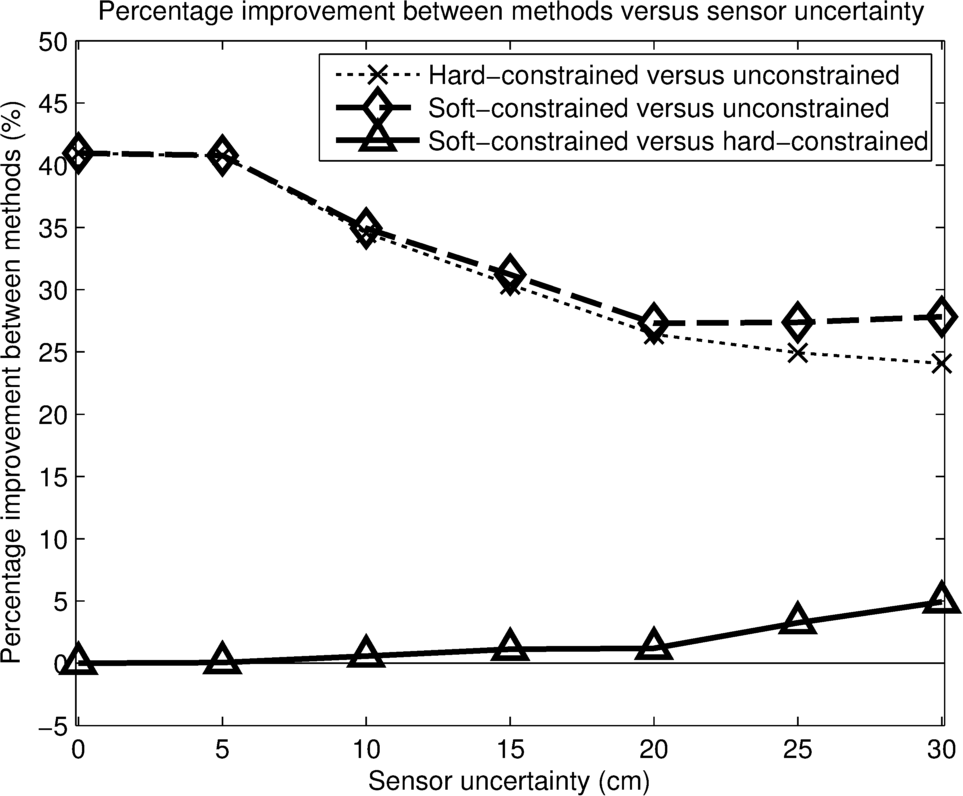}\label{sf:results1_percent}
  }
  \caption{RMSE and percentage improvement between methods for Robot A as the sensor uncertainty is varied. The soft-constrained approach is equal to or better than the unconstrained and hard-constrained approaches in all cases. }
  \label{f:results1}
\end{figure}

\begin{figure}
  \centering
  \subfloat[RMSE]{
    \includegraphics[width=0.6\textwidth]{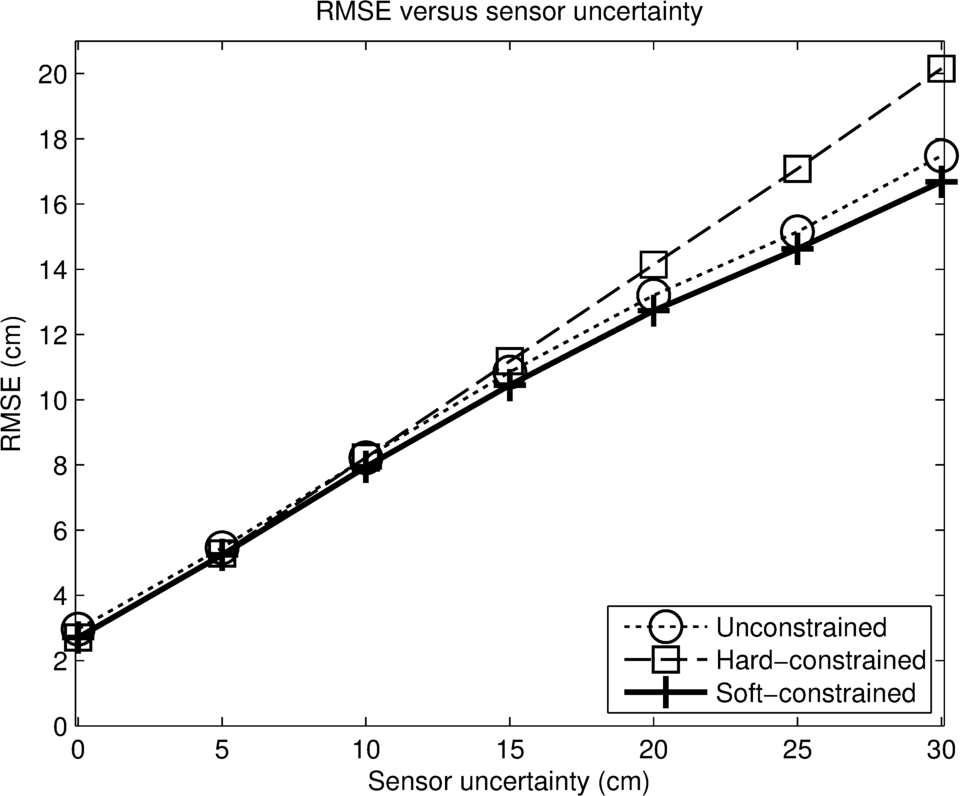}\label{sf:results2_rmse}
  }
  
  \subfloat[Percentage improvement between methods]{
    \includegraphics[width=0.6\textwidth]{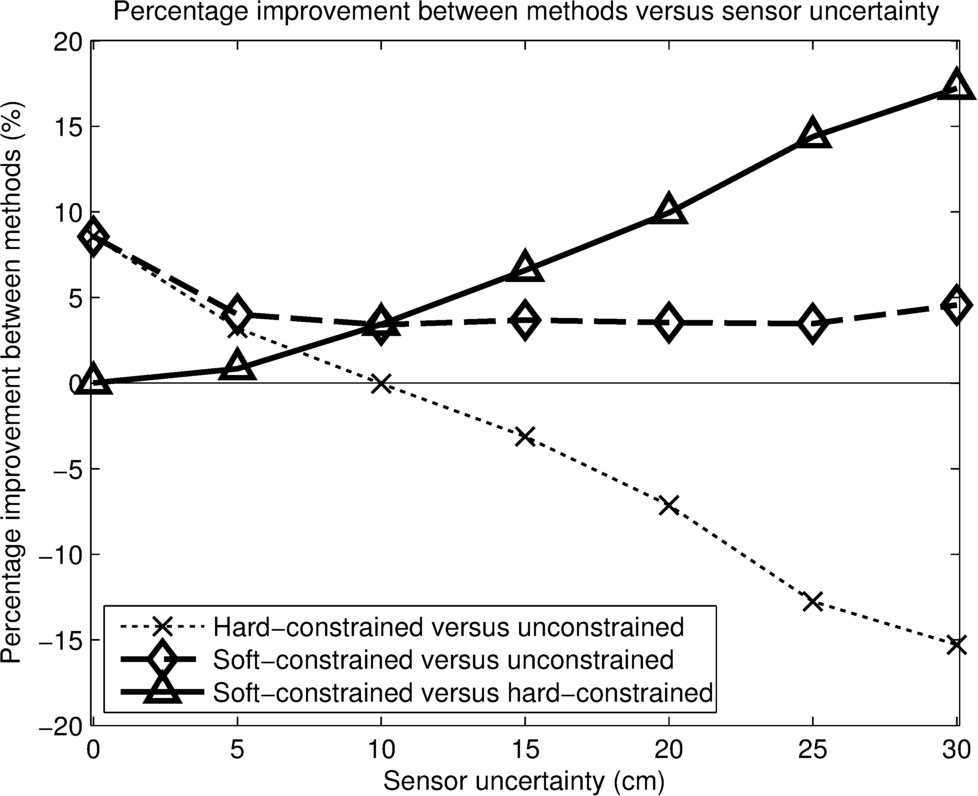}\label{sf:results2_percent}
  }
  \caption{RMSE and percentage improvement between methods for Robot B as the sensor uncertainty is varied. The soft-constrained approach is equal to or better than the unconstrained and hard-constrained approaches in all cases. The hard-constrained approach is outperformed by the unconstrained approach when the sensor uncertainty is large. }
  \label{f:results2}
\end{figure}

\begin{figure}
  \centering
  \subfloat[Unconstrained Kalman filter]{
    \includegraphics[width=0.45\textwidth]{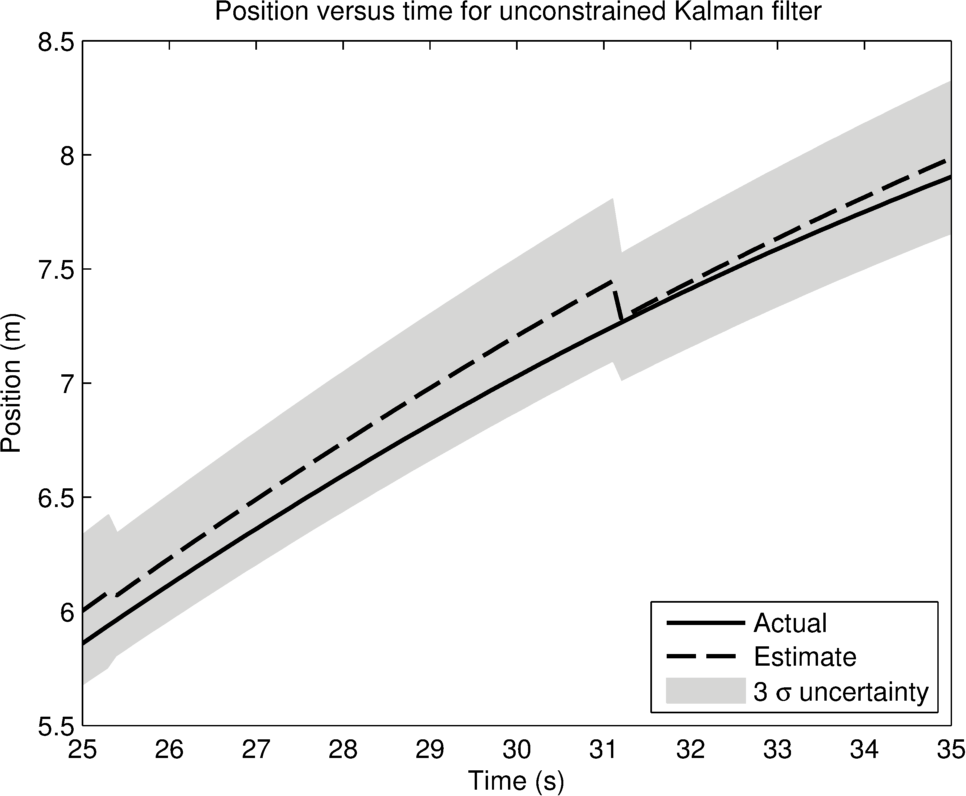}\label{sf:simulation_unc}
  }
  \subfloat[Hard-constrained Kalman filter]{
    \includegraphics[width=0.45\textwidth]{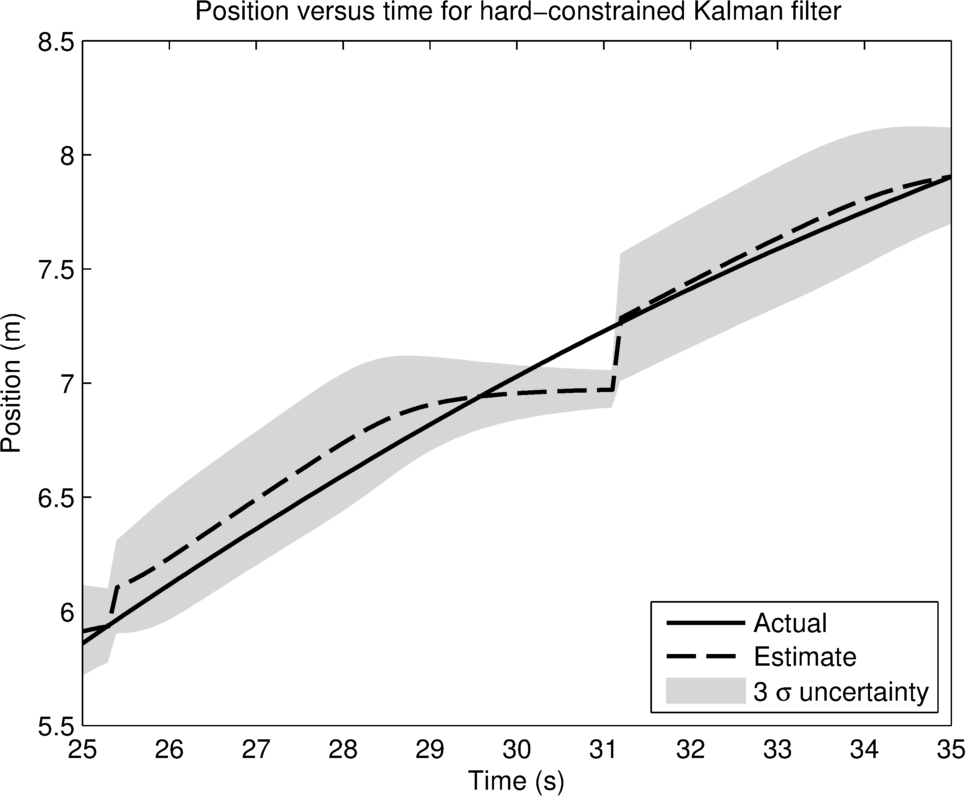}\label{sf:simulation_hard}
  }
  
  \subfloat[Soft-constrained Kalman filter]{
    \includegraphics[width=0.45\textwidth]{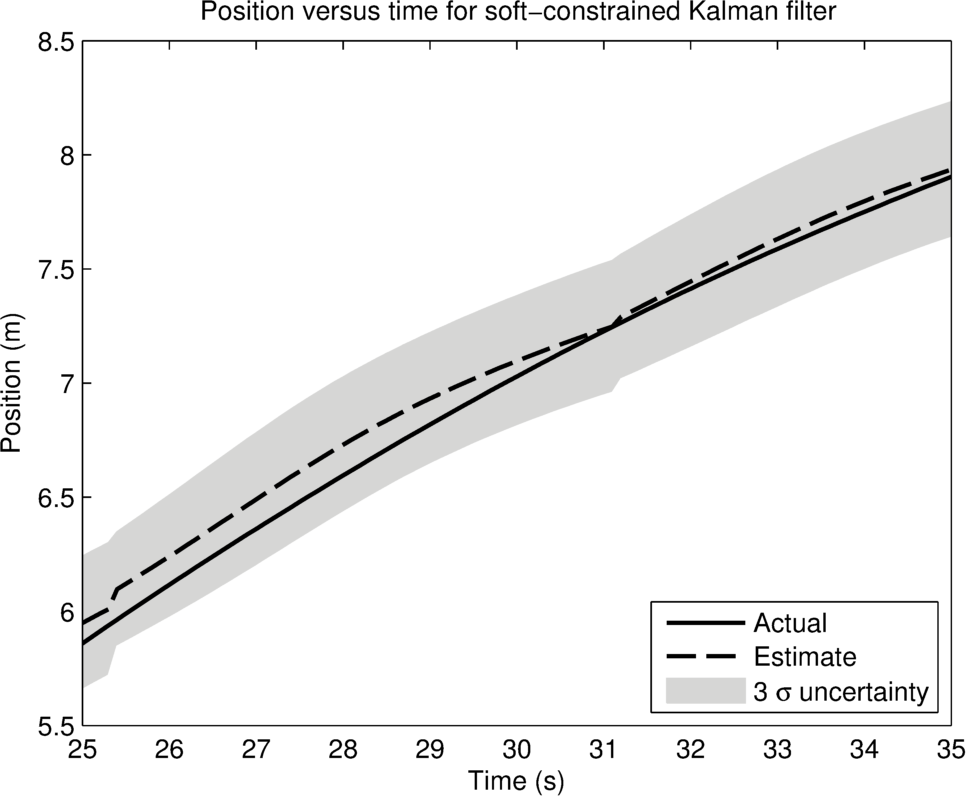}\label{sf:simulation_soft}
  }
  \caption{Comparison of the actual and estimated positions with uncertainty for Robot B with sensor uncertainty of 15cm. This illustrates a case where the hard-constrained approach is outperformed by the unconstrained and soft-constrained approaches. The unconstrained estimate has large uncertainty, while the hard-constrained estimate is overconfident. The soft-constrained approach has a lower uncertainty compared to the unconstrained approach without producing the overconfident estimates of the hard-constrained method. }
  \label{f:simulation_example}
\end{figure}

The time-average Root Mean Square Error (RMSE) and percentage improvement between methods for Robot A are shown in Figure \ref{f:results1}, and the results for Robot B are shown in Figure \ref{f:results2}. For Robot A, both the hard-constrained and soft-constrained methods provided a significant benefit over the unconstrained Kalman filter, with an improvement of over 40\% in tracking performance when the sensors have no uncertainty. As the uncertainty of the sensors was increased, the soft-constrained method slightly outperformed the hard-constrained method. The process noise for Robot B was significantly less than Robot A. As a result, the uncertainty of the sensors played a larger role in determining the performance of the methods. As can be seen in Figure \ref{f:results2}, the hard-constrained Kalman filter was significantly outperformed by the unconstrained Kalman filter once the sensor uncertainty was above 10cm. In these cases, the estimate produced by the hard-constrained Kalman filter was overconfident, and the proposed method outperformed the hard-constrained Kalman filter by over 17\%. An example of the overconfident estimates produced by the hard-constrained method is shown in Figure \ref{f:simulation_example}. The proposed method strikes a balance between the high uncertainty of the unconstrained Kalman filter and the overconfident estimates of the hard-constrained Kalman filter. 

\begin{figure}
  \centering
  \subfloat[Soft-constrained Kalman filter without feedback of the truncated estimate into the Kalman filter]{
    \includegraphics[width=0.6\textwidth]{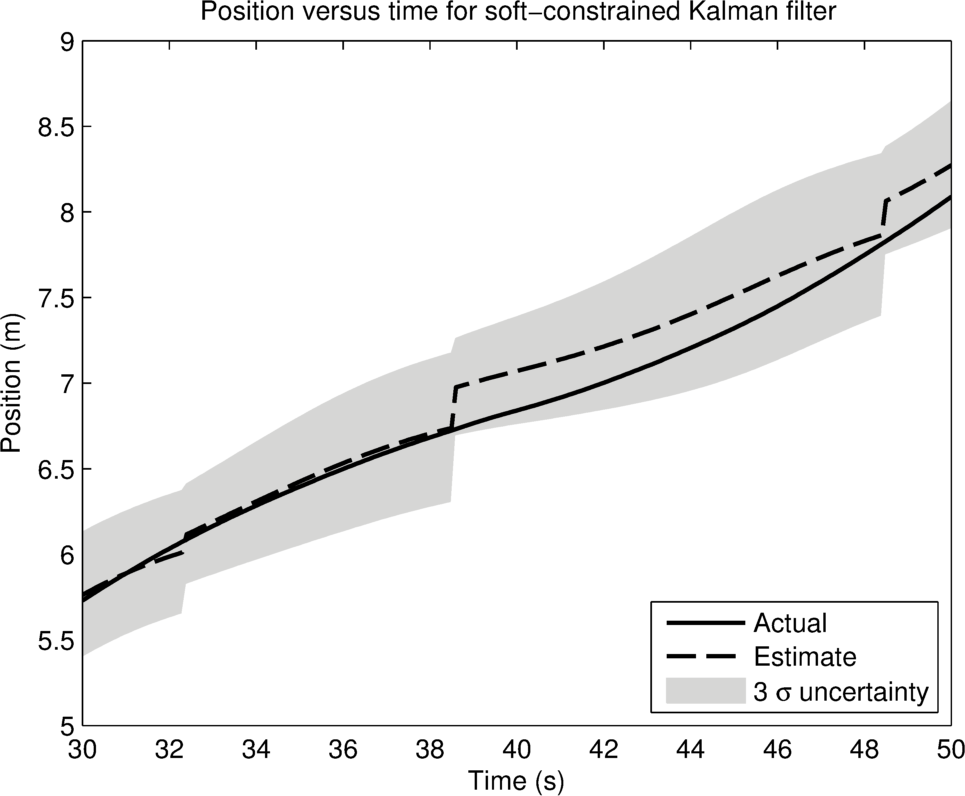}\label{sf:no_feedback}
  }
  
  \subfloat[Soft-constrained Kalman filter with feedback of the truncated estimate into the Kalman filter]{
    \includegraphics[width=0.6\textwidth]{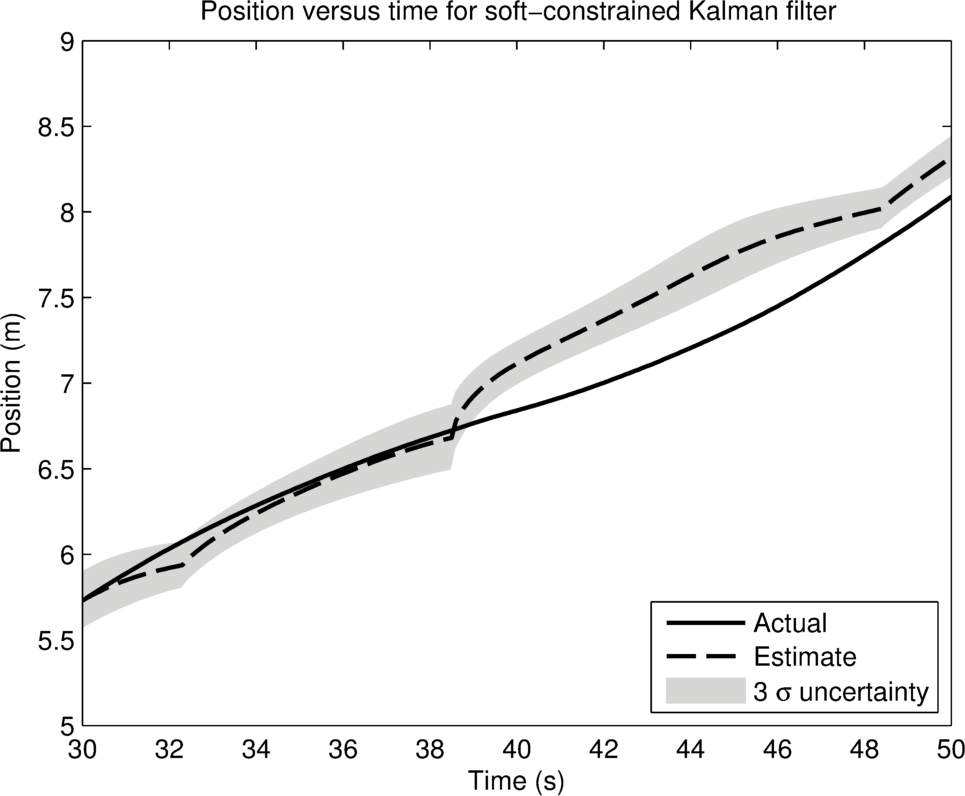}\label{sf:feedback}
  }
  \caption{Comparison of the actual and estimated positions with uncertainty for Robot A with sensor uncertainty of 15cm. In \protect\subref{sf:no_feedback}, the truncated estimate is not fed back into the Kalman filter, while in \protect\subref{sf:feedback}, the truncated estimate is used by the Kalman filter and the resultant estimate is overconfident. }
  \label{f:feedback}
\end{figure}

As discussed at the end of Section \ref{s:constrained}, under certain conditions the truncated state distribution can be fed back into the Kalman filter. For the scenario considered, the discrete position sensors are uncertain rather than noisy, and thus using feedback is not valid. Figure \ref{sf:feedback} shows the effects of using feedback in the example scenario, with the result that the estimate is more confident. However, in some cases this estimate can become overconfident and fail to accurately represent the actual state. The method without feedback (Figure \ref{sf:no_feedback}) has less confident estimates, but they are not overconfident. 

\section{Conclusion}\label{s:conc}

This paper developed an analytical method of truncating an inequality constrained Gaussian distributed variable where the constraints themselves are described by Gaussian distributions. A key aspect of the approach was the use of moment-based Gaussian approximations of the truncated distribution. This truncation method was applied to the constrained Kalman filtering problem where it was shown to outperform unconstrained Kalman filtering and the existing constrained Kalman filter using hard constraints in a simulation example. A key benefit of the developed method compared to hard-constrained Kalman filters is that it is not overconfident near the uncertain constraints. It is an analytical version of existing numerical integration methods, thus providing a computational benefit over the existing numerical methods. 

\section*{Acknowledgements}

This work was supported by the Rio Tinto Centre for Mine Automation and the Australian Centre for Field Robotics, University of Sydney, Australia.

\appendix

\section{Calculation of the mean}\label{s:a_mean}

The integral in equation (\ref{eq:single_mu}) is calculated as follows:

\begin{align}
\int\limits_{-\infty}^{\infty}\zeta &\exp\left(-\zeta^{2}/2\right)\left[1+\textrm{erf}\left(\frac{\zeta-\mu_{c,i}}{\sigma_{c,i}\sqrt{2}} \right)\right] \textrm{d}\zeta \notag \\
&=  \int\limits_{-\infty}^{\infty}\zeta \exp\left(-\zeta^{2}/2\right) \textrm{d}\zeta \label{eq:a_mean_first_int}\\
&\quad + \int\limits_{-\infty}^{\infty}\zeta \exp\left(-\zeta^{2}/2\right)\textrm{erf}\left(\frac{\zeta-\mu_{c,i}}{\sigma_{c,i}\sqrt{2}} \right) \textrm{d}\zeta \label{eq:a_mean_second_int}
\end{align}

The first integral, (\ref{eq:a_mean_first_int}), equates to 0. For integral (\ref{eq:a_mean_second_int}), performing integration by parts with:
\begin{equation}
u = \textrm{erf}\left(\frac{\zeta - \mu_{c,i}}{\sigma_{c,i}\sqrt{2}} \right) \qquad \textrm{d}v = \zeta \exp\left(-\zeta^{2}/2\right)\textrm{d}\zeta
\end{equation}
gives:
\begin{align}
\int\limits_{-\infty}^{\infty}&\zeta \exp\left(-\zeta^{2}/2\right)\textrm{erf}\left(\frac{\zeta-\mu_{c,i}}{\sigma_{c,i}\sqrt{2}} \right) \textrm{d}\zeta \notag \\
& = \left[-\exp\left(-\zeta^{2}/2\right) \textrm{erf}\left(\frac{\zeta - \mu_{c,i}}{\sigma_{c,i}\sqrt{2}} \right)  \right]^{\infty}_{-\infty} \label{eq:a4_int} \\
& \quad + \int\limits_{-\infty}^{\infty} \frac{\exp\left(-\zeta^{2}/2\right)}{\sigma_{c,i}}\sqrt{\frac{2}{\pi}}\exp\left(-\frac{(\zeta-\mu_{c,i})^{2}}{2\sigma_{c,i}^{2}}\right) \textrm{d}\zeta \label{eq:a5_int}
\end{align}

Equation (\ref{eq:a4_int}) equals 0, and completing the square allows (\ref{eq:a5_int}) to be calculated as:

\begin{equation} \label{eq:single_mu_completed_square}
\int\limits_{-\infty}^{\infty} \frac{\exp\left(-\zeta^{2}/2\right)}{\sigma_{c,i}}\sqrt{\frac{2}{\pi}}\exp\left(-\frac{(\zeta-\mu_{c,i})^{2}}{2\sigma_{c,i}^{2}}\right) \textrm{d}\zeta \\
= \frac{2\exp\left(-\frac{\mu_{c,i}^{2}}{2(\sigma_{c,i}^{2}+1)}\right)}{\sqrt{\sigma_{c,i}^{2}+1}}
\end{equation}

To summarise:

\begin{equation} 
\boxed{
\int\limits_{-\infty}^{\infty}\zeta \exp\left(-\zeta^{2}/2\right)\left[1+\textrm{erf}\left(\frac{\zeta-\mu_{c,i}}{\sigma_{c,i}\sqrt{2}} \right)\right] \textrm{d}\zeta    = \frac{2}{\sqrt{\sigma_{c,i}^2 + 1}}\exp\left(-\frac{\mu_{c,i}^{2}}{2(\sigma_{c,i}^2 + 1)}\right) \label{eq:a_single_mu}
}
\end{equation}

\section{Calculation of the variance}\label{s:a_variance}

The integral in equation (\ref{eq:single_sigma}) is calculated as follows:

\begin{align}
\int\limits_{-\infty}^{\infty}\left(\zeta - \mu_{i}\right)^{2} &\exp\left(-\zeta^{2}/2\right)\left[1+\textrm{erf}\left(\frac{\zeta-\mu_{c,i}}{\sigma_{c,i}\sqrt{2}} \right)\right] \textrm{d}\zeta \notag\\ 
&= \int\limits_{-\infty}^{\infty}\zeta^{2}\exp\left(-\zeta^{2}/2\right) \textrm{d}\zeta \label{eq:single_sigma1} \\
&\quad + \int\limits_{-\infty}^{\infty}\zeta^{2}\exp\left(-\zeta^{2}/2\right)\textrm{erf}\left(\frac{\zeta-\mu_{c,i}}{\sigma_{c,i}\sqrt{2}} \right) \textrm{d}\zeta \label{eq:single_sigma2}\\
&\quad - \int\limits_{-\infty}^{\infty} 2\mu_{i}\zeta \exp\left(-\zeta^{2}/2\right) \textrm{d}\zeta \label{eq:single_sigma3}\\ 
&\quad - \int\limits_{-\infty}^{\infty}2\mu_{i}\zeta \exp\left(-\zeta^{2}/2\right)\textrm{erf}\left(\frac{\zeta-\mu_{c,i}}{\sigma_{c,i}\sqrt{2}} \right) \textrm{d}\zeta \label{eq:single_sigma4}\\
&\quad + \int\limits_{-\infty}^{\infty}\mu_{i}^{2} \exp\left(-\zeta^{2}/2\right) \textrm{d}\zeta \label{eq:single_sigma5} \\
&\quad + \int\limits_{-\infty}^{\infty}\mu_{i}^{2} \exp\left(-\zeta^{2}/2\right)\textrm{erf}\left(\frac{\zeta-\mu_{c,i}}{\sigma_{c,i}\sqrt{2}} \right) \textrm{d}\zeta \label{eq:single_sigma6}
\end{align}

Integral (\ref{eq:single_sigma1}) equates to $\sqrt{2\pi}$. For integral (\ref{eq:single_sigma2}), performing integration by parts with:
\begin{equation}
u = \textrm{erf}\left(\frac{\zeta - \mu_{c,i}}{\sigma_{c,i}\sqrt{2}} \right) \qquad \textrm{d}v = \zeta^{2} \exp\left(-\zeta^{2}/2\right)\textrm{d}\zeta
\end{equation}
gives:

\begin{align}
&\int\limits_{-\infty}^{\infty}\zeta^{2}\exp\left(-\zeta^{2}/2\right)\textrm{erf}\left(\frac{\zeta-\mu_{c,i}}{\sigma_{c,i}\sqrt{2}} \right) \textrm{d}\zeta \notag \\
&= \left[ \textrm{erf}\left(\frac{\zeta - \mu_{c,i}}{\sigma_{c,i}\sqrt{2}} \right)\left(\sqrt{\frac{\pi}{2}}\textrm{erf}\left(\frac{\zeta}{\sqrt{2}} \right)-\zeta \exp\left(-\zeta^{2}/2\right) \right) \right]^{\infty}_{-\infty} \label{eq:single_sigma_2_part1}\\
&\quad - \int\limits_{-\infty}^{\infty} \frac{1}{\sigma_{c,i}}\sqrt{\frac{2}{\pi}}\exp\left(-\frac{(\zeta-\mu_{c,i})^{2}}{2\sigma_{c,i}^{2}}\right)\left(\sqrt{\frac{\pi}{2}}\textrm{erf}\left(\frac{\zeta}{\sqrt{2}} \right)-\zeta \exp\left(-\zeta^{2}/2\right)\right) \textrm{d}\zeta \label{eq:single_sigma_2_part2}
\end{align}

Equation (\ref{eq:single_sigma_2_part1}) is equal to 0, and integral (\ref{eq:single_sigma_2_part2}) can be split into the following integrals:
\begin{align}
\int\limits_{-\infty}^{\infty} \frac{1}{\sigma_{c,i}}\sqrt{\frac{2}{\pi}}&\exp\left(-\frac{(\zeta-\mu_{c,i})^{2}}{2\sigma_{c,i}^{2}}\right)\left(\sqrt{\frac{\pi}{2}}\textrm{erf}\left(\frac{\zeta}{\sqrt{2}} \right)-\zeta \exp\left(-\zeta^{2}/2\right)\right) \textrm{d}\zeta \notag \\
&= \int\limits_{-\infty}^{\infty} \frac{\exp\left(-\frac{(\zeta-\mu_{c,i})^{2}}{2\sigma_{c,i}^{2}}\right)}{\sigma_{c,i}}\textrm{erf}\left(\frac{\zeta}{\sqrt{2}} \right) \textrm{d}\zeta \label{eq:single_sigma_2_part2_1} \\
&\quad- \int\limits_{-\infty}^{\infty} \sqrt{\frac{2}{\pi}}\frac{\zeta}{\sigma_{c,i}}\exp\left(-\frac{(\zeta-\mu_{c,i})^{2}}{2\sigma_{c,i}^{2}}\right)\exp\left(-\zeta^{2}/2\right) \textrm{d}\zeta \label{eq:single_sigma_2_part2_2}
\end{align}

Integral (\ref{eq:single_sigma_2_part2_1}) does not have an indefinite integral, but a definite integral is provided in \cite{Fayed2014}:

\begin{equation}
\int\limits_{-\infty}^{\infty} \exp\left(-(\alpha t+\beta)^{2}\right) \textrm{erf}\left(at+b\right) \textrm{d}t = \frac{\sqrt{\pi}}{\alpha}\textrm{erf}\left[\frac{\alpha b - \beta a}{\sqrt{\alpha^{2} + a^{2}}}\right]
\end{equation}

Using this formula, integral (\ref{eq:single_sigma_2_part2_1}) equates to:
\begin{equation}
\int\limits_{-\infty}^{\infty} \frac{\exp\left(-\frac{(\zeta-\mu_{c,i})^{2}}{2\sigma_{c,i}^{2}}\right)}{\sigma_{c,i}}\textrm{erf}\left(\frac{\zeta}{\sqrt{2}} \right) \textrm{d}\zeta = \sqrt{2\pi}\textrm{erf}\left(\frac{\mu_{c,i}}{\sqrt{2(\sigma_{c,i}^{2} + 1)}} \right)
\end{equation}

Integral (\ref{eq:single_sigma_2_part2_2}) has the same exponent as in equation (\ref{eq:single_mu_completed_square}). Completing the square gives:

\begin{equation}
\int\limits_{-\infty}^{\infty} \sqrt{\frac{2}{\pi}}\frac{\zeta}{\sigma_{c,i}}\exp\left(-\frac{(\zeta-\mu_{c,i})^{2}}{2\sigma_{c,i}^{2}}\right)\exp\left(-\zeta^{2}/2\right) \textrm{d}\zeta = \frac{2\mu_{c,i}\exp\left(-\frac{\mu_{c,i}^{2}}{2(\sigma_{c,i}^{2}+1)}\right)}{(\sigma_{c,i}^{2} + 1)^{3/2}}
\end{equation}

To summarise, integral (\ref{eq:single_sigma2}) simplifies to:

\begin{multline}
\int\limits_{-\infty}^{\infty}\zeta^{2}\exp\left(-\zeta^{2}/2\right)\textrm{erf}\left(\frac{\zeta-\mu_{c,i}}{\sigma_{c,i}\sqrt{2}} \right) \textrm{d}\zeta \\
= \frac{2\mu_{c,i}\exp\left(-\frac{\mu_{c,i}^{2}}{2(\sigma_{c,i}^{2}+1)}\right)}{(\sigma_{c,i}^{2} + 1)^{3/2}} -\sqrt{2\pi}\textrm{erf}\left(\frac{\mu_{c,i}}{\sqrt{2(\sigma_{c,i}^{2} + 1)}} \right)
\end{multline}

Integral (\ref{eq:single_sigma3}) equates to 0. Integral (\ref{eq:single_sigma4}) is similar to the integral evaluated in \ref{s:a_mean} and equates to:

\begin{equation}
\int\limits_{-\infty}^{\infty}2\mu_{i}\zeta \exp\left(-\zeta^{2}/2\right)\textrm{erf}\left(\frac{\zeta-\mu_{c,i}}{\sigma_{c,i}\sqrt{2}} \right) \textrm{d}\zeta = \frac{4\mu_{i}}{\sqrt{\sigma_{c,i}^{2} + 1}} \exp\left(-\frac{\mu_{c,i}^{2}}{\sqrt{\sigma_{c,i}^{2} + 1}}\right)
\end{equation}

Integral (\ref{eq:single_sigma5}) is the integral of a Gaussian distribution and equates to $\mu_{i}^{2}\sqrt{2\pi}$. Using the formula provided in \cite{Fayed2014}, integral (\ref{eq:single_sigma6}) equates to:

\begin{equation}
\int\limits_{-\infty}^{\infty}\mu_{i}^{2} \exp\left(-\zeta^{2}/2\right)\textrm{erf}\left(\frac{\zeta-\mu_{c,i}}{\sigma_{c,i}\sqrt{2}} \right) \textrm{d}\zeta = \mu_{i}^{2}\sqrt{2\pi}\textrm{erf}\left(\frac{-\mu_{c,i}}{\sqrt{2(\sigma_{c,i}^{2} + 1)}}  \right)
\end{equation}

Summarising:

\begin{equation} \label{eq:a_single_sigma_summary}
\boxed{
\begin{aligned}
\int\limits_{-\infty}^{\infty}\left(\zeta - \mu_{i}\right)^{2} &\exp\left(-\zeta^{2}/2\right)\left[1+\textrm{erf}\left(\frac{\zeta-\mu_{c,i}}{\sigma_{c,i}\sqrt{2}} \right)\right] \textrm{d}\zeta    \\
&= \left[\sqrt{2\pi}\left(\left(1+\mu_{i}^{2}\right)\left(1-\textrm{erf}\left(\frac{\mu_{c,i}}{\sqrt{2(\sigma_{c,i}^{2}+1)}}\right)\right)\right) \right. \\
& \quad + \left. \frac{2}{\sqrt{\sigma_{c,i}^2 + 1}}\exp\left(-\frac{\mu_{c,i}^{2}}{2(\sigma_{c,i}^2 + 1)}\right)\left(\frac{\mu_{c,i}}{\sigma_{c,i}^2 + 1} - 2\mu_{i} \right)\right]
\end{aligned}
}
\end{equation}



\bibliographystyle{elsarticle-num} 
\bibliography{bib}


%
%
%
\end{document}